\documentclass{aa}
\usepackage{natbib}
\usepackage{txfonts}
\usepackage{graphicx}

\begin{document}

\title{Three years in the coronal life of AB Dor}
\subtitle{I. Plasma emission measure distributions and abundances \\ at
  different activity levels.} 
\author{J. Sanz-Forcada, A. Maggio, \and G. Micela}
\institute{INAF - Osservatorio Astronomico di Palermo
G. S. Vaiana, Piazza del Parlamento, 1; Palermo, I-90134, Italy}
\offprints{J. Sanz-Forcada, \email{jsanz@astropa.unipa.it}}
\date{Received / Accepted}

\abstract{
The young active star AB~Dor (K1 IV-V) has been observed 16 times in
the last three years with the XMM-Newton and Chandra observatories,
totalling 650~ks of high-resolution X-ray spectra. The XMM/RGS
observations with the highest and lowest average 
emission levels have been selected to study
the coronal properties of AB~Dor in two different activity levels. 
We compare the results based on the XMM data with those obtained from
a higher resolution Chandra/HETG spectrum, using the same line-based 
analysis technique. We have reconstructed the plasma
Emission Measure Distribution vs.\ temperature (EMD) in the range
log~$T$(K)$\sim$6.1--7.6, and we have determined the coronal abundances of
AB~Dor, obtaining consistent results between the two instruments.
The overall shape of the EMD is also consistent with the one previously
inferred from EUVE data.
The EMD shows a steep increase up to the peak at log~$T$(K)$\sim$6.9 and
a substantial amount of plasma in the range 
log~$T$(K)$\sim$6.9--7.3. The
coronal abundances show a clear trend of 
increasing depletion with respect to solar photospheric values, for
elements with increasing First Ionization Potential (FIP), down to the Fe
value ([Fe/H] = --0.57), followed by a more gradual recovery of the
photospheric values for elements with higher FIP.
He-like triplets and \ion{Fe}{xxi}
and \ion{Fe}{xxii} lines ratios indicate electron densities log~$n_{\rm e}
\sim 10.8$\,cm$^{-3}$ at log~$T$(K)$\sim 6.3$ and log~$n_{\rm
e}\sim 12.5$\,cm$^{-3}$ at log~$T$(K)$\sim$7, implying plasma
pressures steeply increasing with temperature. These results are
interpreted in the framework of a corona composed of different families
of magnetic loop structures, shorter than the stellar radius and in
isobaric conditions, having pressures increasing with the maximum plasma
temperature, and which occupy a small fraction ($f \sim
10^{-4}$--$10^{-6}$) of the stellar surface.

}

   \maketitle
%
\section{Introduction}

\object{AB Dor} (HD 36705, K1 IV-V) is a frequent target for
studies on stellar activity.  Almost 
arrived in the main sequence, with an age of $\sim$20--30~Myr 
\citep{col97}, it has a high rotation rate 
($P_{\rm rot}$=0.5148 d), and persistent large-scale magnetic field patterns in
its photosphere \citep[see][and references therein]{don99,hus02}. 
Two companions, not expected to interact
with AB Dor, have been detected in the vicinity. The dM4e star Rossiter
137B (AB~Dor B), detected as a faint source also in X-rays \citep{vil87}, is
10$\arcsec$ away from the main source, while the third companion is a
very low-mass star (0.08--0.11 M$_{\sun}$) at a distance of
0.2$\arcsec$--0.7$\arcsec$ (3--10 a.u.) from the primary
\citep{gui97}. The contribution of the
companions to the X-ray spectrum of the main source can be considered
negligible, essentially because the quiescent X-ray emission of the
companions scales as their bolometric luminosity (in fact, for their
young age, all the stars in the system emit close to the saturation level of
$L_{\rm X}$/$L_{\rm bol}\sim$10$^{-3}$).
AB~Dor has been observed with the main space observatories in the UV, EUV,
and X-rays, like 
HST, FUSE, ROSAT, BeppoSAX, ASCA and EUVE \citep[see][and references
therein]{vil98,ake00,sch98,kue97,mag00,ruc95,mew96,sanz02},
showing 
rotational modulation in some lines formed in the transition region,
and a corona dominated by material at temperatures of
log~$T$(K)$\sim$6.7--7.3, significantly higher than in the solar 
quiescent corona. 
Most recently, initial results from the first XMM observations of AB~Dor, 
taken in May and June 2000, have been presented by \citet{gud01},
while a preliminary analysis of the first Chandra observation was made by
\citet{lin01}.

\begin{figure}
\includegraphics[width=0.33\textwidth,angle=90]{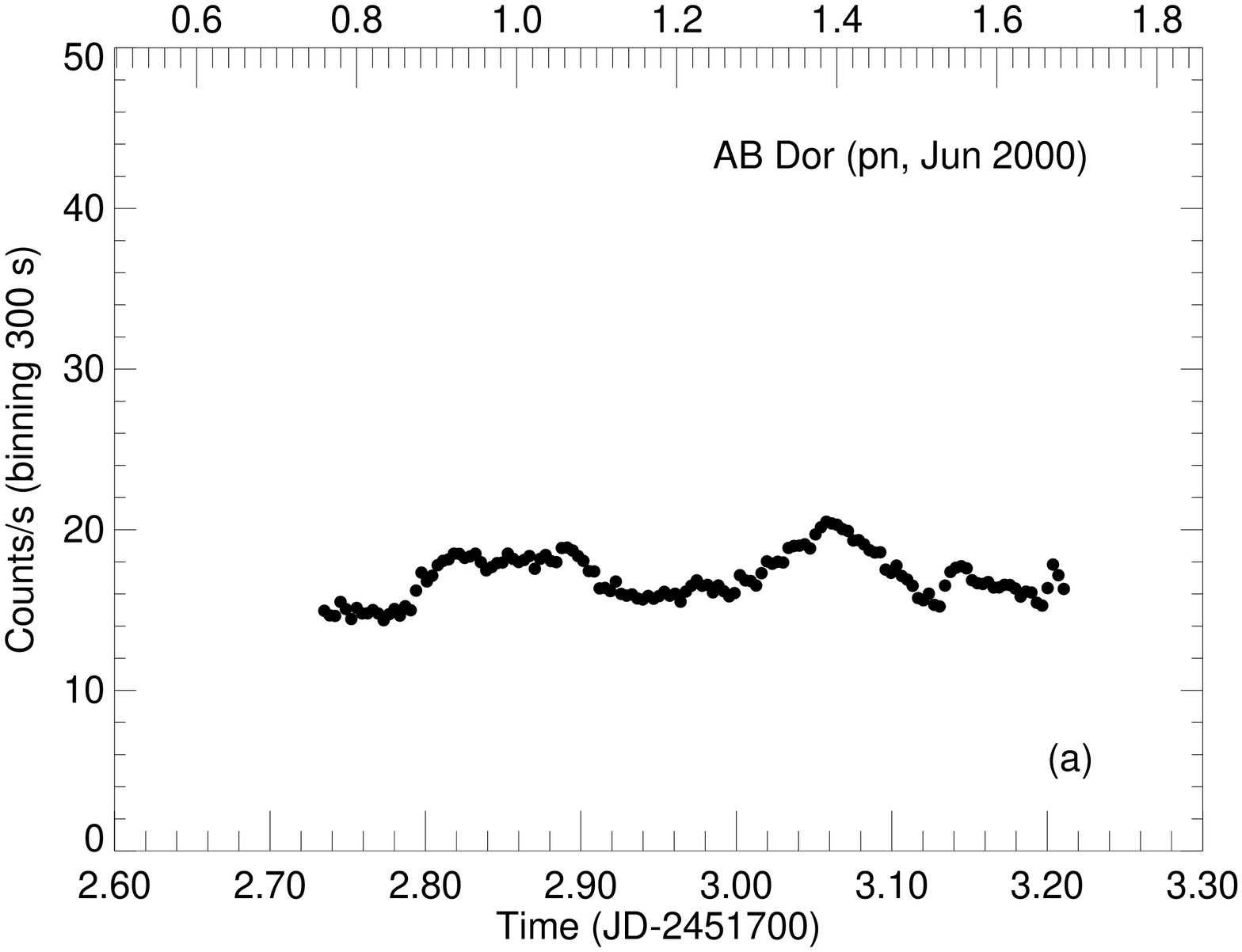}
\includegraphics[width=0.33\textwidth,angle=90]{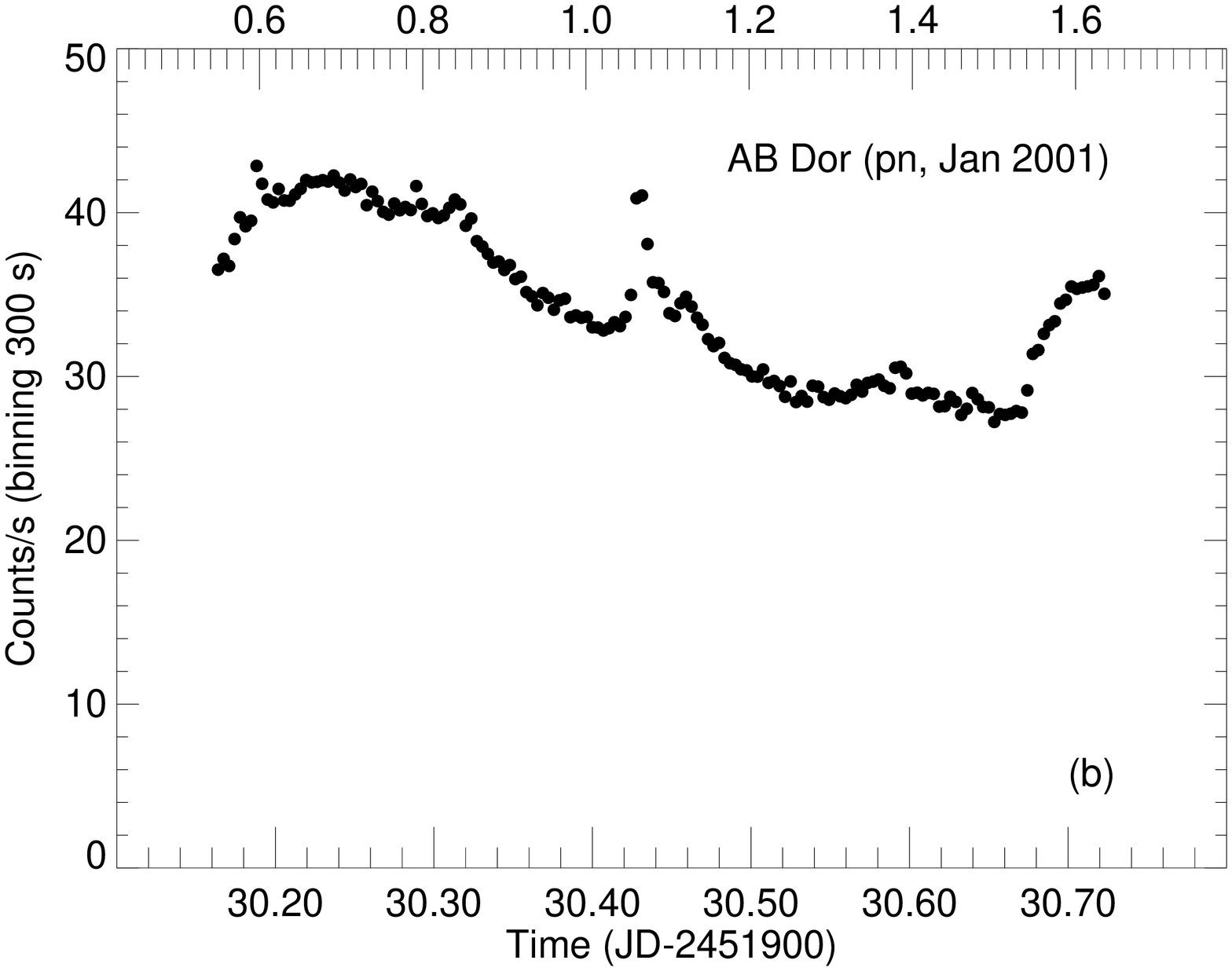}
\includegraphics[width=0.33\textwidth,angle=90]{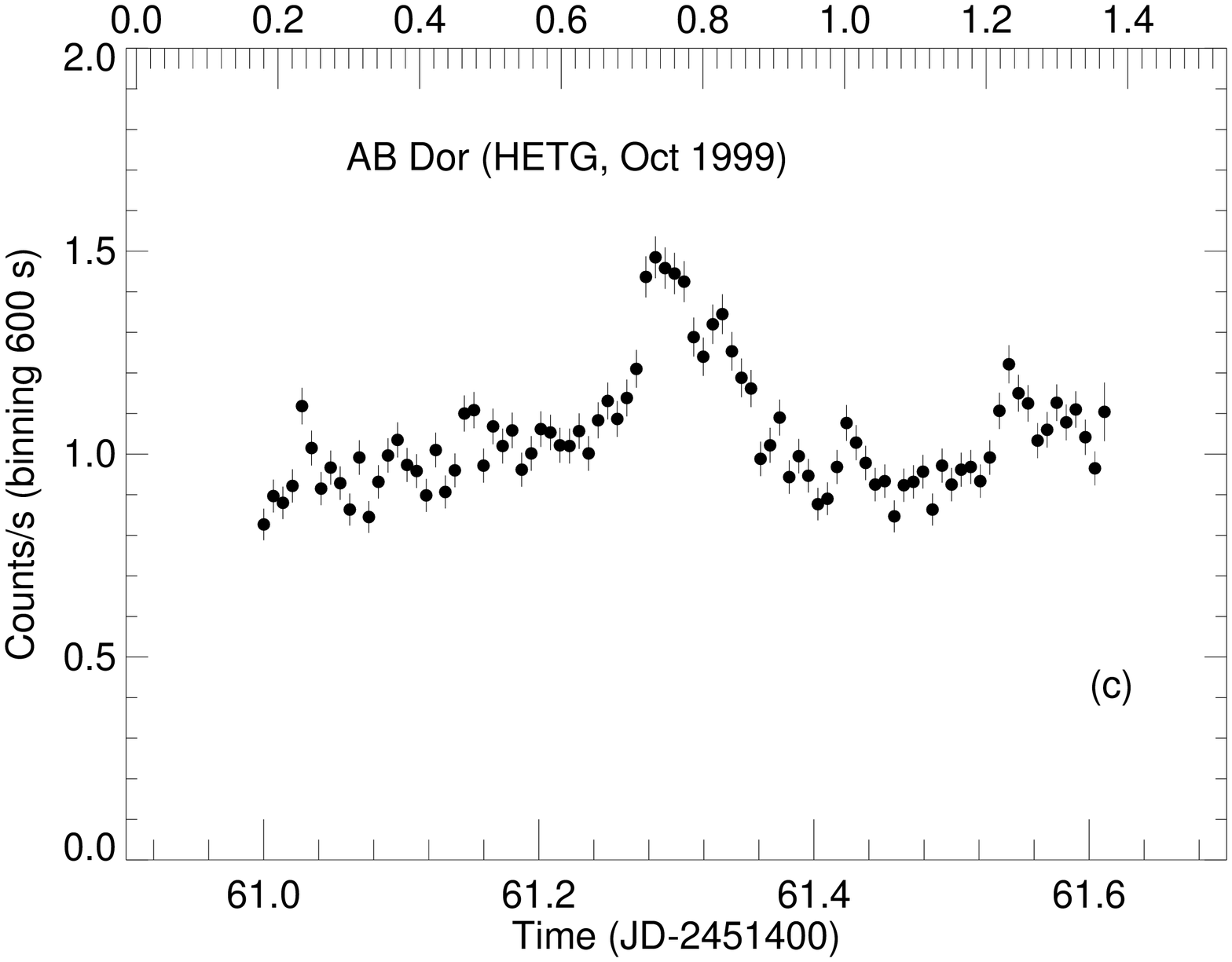}
 \caption{X-ray light curves from the observations analyzed
   in this work:
   {($a$)} and {($b$)}: XMM/EPIC-pn light curves of rev.~\#091 and \#205;
   {($c$)}: Chandra/HETG 1st order light curve of AB Dor A. 
   The lower axis indicates time in days, and the upper axis
   reports the rotational phase, using ephemeris by
   \citet{inn88}. 
   Variations of 35--50\% in the flux level are found during each of 
   the three observations.\label{lightcurves}}
\end{figure}

Rotational modulation up to a factor $\sim 2$ has been detected in \ion{C}{iii}
and \ion{O}{vi} lines which
form in the upper chromosphere and in the transition region (Ake et al.
2000), but only at about the 15\% level for the X-ray emission observed
with ROSAT (K\"urster et al. 1997).
Several EUV and X-ray spectroscopic studies show that the
corona of AB~Dor is dominated by plasma at temperatures of
log~$T$(K)$\sim$6.7--7.3, significantly higher than in the solar 
quiescent corona. On the other hand, only few determinations of the plasma 
density have been published up to date: using density-sensitive \ion{C}{iii}
lines in {\sc orpheus} and {\sc fuse} spectra, densities
of $\sim 10^{11}$\,cm$^{-3}$ or more at $T \sim 80000$\,K have been determined 
by Schmitt et al. (1998) and by Ake et al. (2000), while G\"udel et al.
(2001) have estimated a coronal density of $3 \times 10^{10}$\,cm$^{-3}$
at $T \sim 2 \times 10^6$\,K using the He-like
\ion{O}{vii} emission line triplet; finally, densities of the order of
$10^{12}$\,cm$^{-3}$ at $T \sim 10^7$\,K have been recently determined by 
Sanz-Forcada et al. (2002) from EUVE spectra.

There are still several open issues concerning the structure of the
corona of AB~Dor, and more in general on the coronae of very active
stars in saturated X-ray emission regime. The first question is whether
the hot coronal plasma is homogeneously distributed across the stellar
surface, or rather it is spatially concentrated. Up to date, limited
and sometimes contradicting information on the sizes and location of the
X-ray emitting coronal structures has been derived from the analysis
and modeling of X-ray flares observed with EXOSAT \citep{col88},
BeppoSAX \citep{mag00}, and XMM-Newton \citep{gud01}, as 
well as from the reconstruction of the three-dimensional 
magnetic field geometry based on Zeeman-Doppler maps
\citep{hus02,jar02}.
A second related question is whether the
coronal emission originates from compact (high-density) structures,
possibly located above the high-latitude ($> 60^{\degr}$) spots 
suggested by Doppler images \citep[][and references therein]{don99},
or perhaps also from the large-scale structures suggested to
explain the stable slingshot prominences revealed by transient
absorption features in the H$\alpha$ line \citep{col89}. 
The above questions, and the related issues on nature of the
magnetic dynamo activity in AB~Dor, can be usefully addressed by
new accurate determinations of the plasma density from spectroscopic
diagnostics and from a detailed study of the X-ray emission
variability of this coronal source.

\begin{figure*}
 \resizebox{\hsize}{!}{\includegraphics[angle=270]{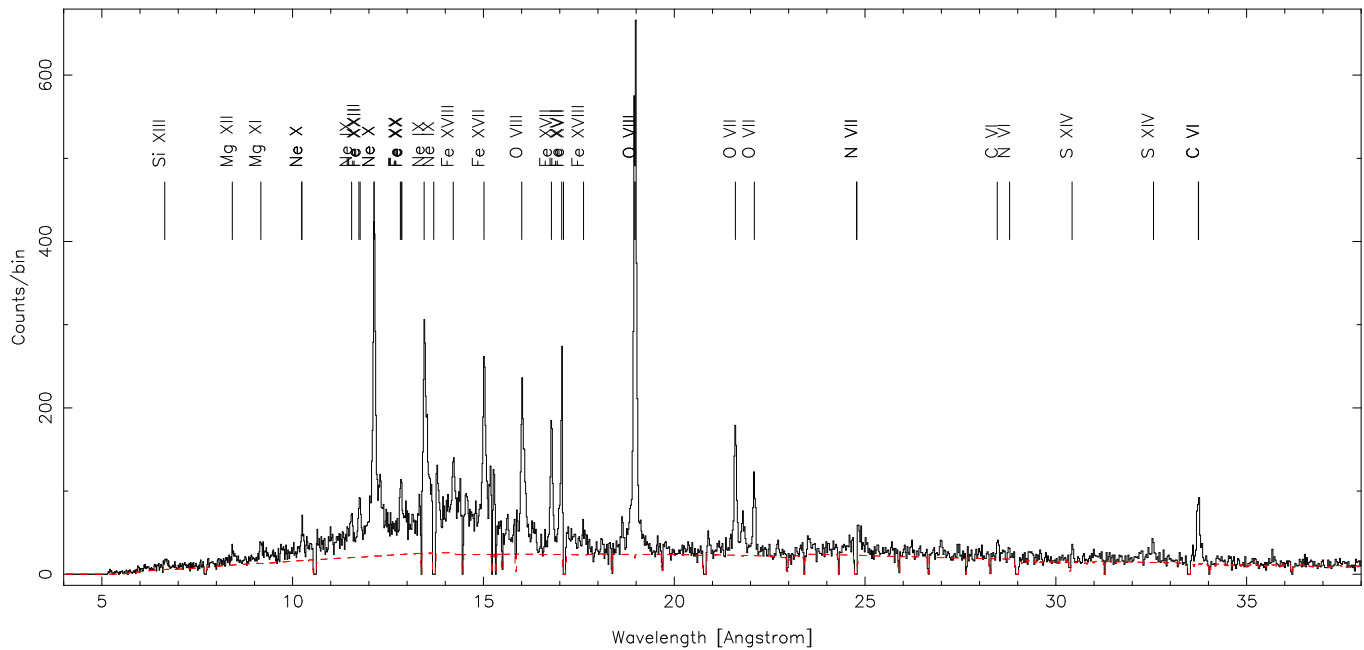}}
 \caption{RGS 1 first order spectrum of AB Dor from the
 revolution \#091 observation. The dashed 
 line represents the continuum predicted by the EMD. A false continuum
 is created by the extended instrumental line profiles.\label{xmmspec}}
\end{figure*}

As one of the brightest X-ray stellar sources, AB Dor has been chosen
for the XMM-Newton calibration program, and 15 observations are
available to date, totalling 594~ks of clean RGS spectra. 
This is the first of a series of papers devoted to a detailed and systematic 
analysis of the XMM-Newton observations available to date, and it is
dedicated to the high resolution
spectra with the lowest and highest global count rates, with the aim
of understanding the properties of the corona of AB~Dor in two
different activity levels. In this paper
we also present the result of a new analysis of the higher resolution
Chandra/HETG spectrum, performed with the same method employed for the
XMM data. 
Issues related to the analysis of XMM/RGS and
Chandra/HETG spectra are discussed, and a comparison with the results
obtained from previous EUVE observations is presented.

The technical information related to the observations is given in
Sect.~2. The methods employed to analyze the data are delineated in
Sect.~\ref{sec:analysis}, as well as the issues that may affect the
measurements in this kind of spectra. The scientific results are
discussed in Sect.~\ref{sec:results}, followed by a summary of
the conclusions in Sect.~\ref{sec:conclusions}.

\section{Observations}

\subsection{XMM-Newton}  

\begin{figure}
 \resizebox{\hsize}{!}{\includegraphics[angle=270]{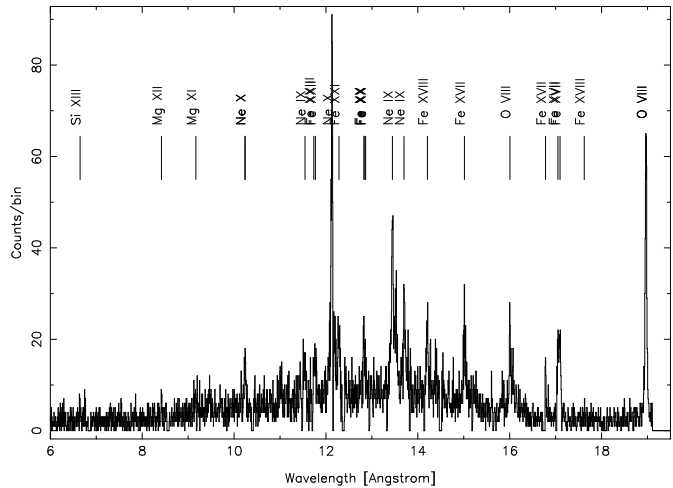}}
 \caption{RGS 1 second order spectrum of AB Dor from
   the rev.~\#091 observation.\label{xmmsecond}}
\end{figure}

AB Dor has been frequently observed as XMM-Newton
calibration target since May 2000 
(Table~\ref{times}), with different combinations of instruments
operating simultaneously. 
XMM-Newton allows to carry out simultaneous observation
with the EPIC (European Imaging Photon Camera) PN and MOS detectors
(sensitivity range 0.15--15 keV and 0.2--10~keV respectively), and
with the RGS \citep[Reflection Grating Spectrometer,][]{denher01} 
($\lambda\lambda\sim$6--38~\AA), 
allowing us to obtain simultaneously medium-resolution CCD spectra
($\Delta E\sim$70~eV at $E\sim$1~keV) 
and high-resolution grating ($\lambda$/$\Delta\lambda\sim$100--500)
spectra. 
The data have been reduced by employing the 
standard tasks present in the SAS (Science Analysis Software) package
v5.3.3, removing the 
time intervals when the background was higher than 0.5 cts/s in CCD
\#9,  in order to ensure a ``clean'' spectrum.  
The average count rates obtained in the RGS 2 spectra, after the
high-background removal, are shown in
Table~\ref{times}. The observations with the lowest and highest RGS count
rates (rev. \#091 and rev. \#205) are analyzed in detail in this work
in order to investigate the differences in the coronal thermal
structure among these two levels of activity.
Light curves for the two observations (Fig.~\ref{lightcurves}a and b) 
were obtained by selecting a circle centered on the source in
the EPIC-pn images, and subtracting the background count rate
taken proportionally.
The image presents a clear asymmetry in the main source, that we
attribute to AB~Dor B; this source can contaminate both
the light curve and the RGS spectra of AB Dor, but we can safely neglect its
effects in the RGS spectra, given the flux ratio observed with Chandra
(see below).
High resolution spectra corresponding to the first order of the RGS
(Fig.~\ref{xmmspec}), together with some of the lines identified,
are analyzed as explained in Sect.~\ref{sec:analysis}. Second order 
RGS spectra have been used to double check for the blends
contributing to the main lines (Fig.~\ref{xmmsecond}), and to test the
flux calibration with respect to the first order (see Sect.~\ref{sec:emdmethod}).

\subsection{Chandra}   

\begin{figure*}
 \resizebox{\hsize}{!}{\includegraphics[angle=270]{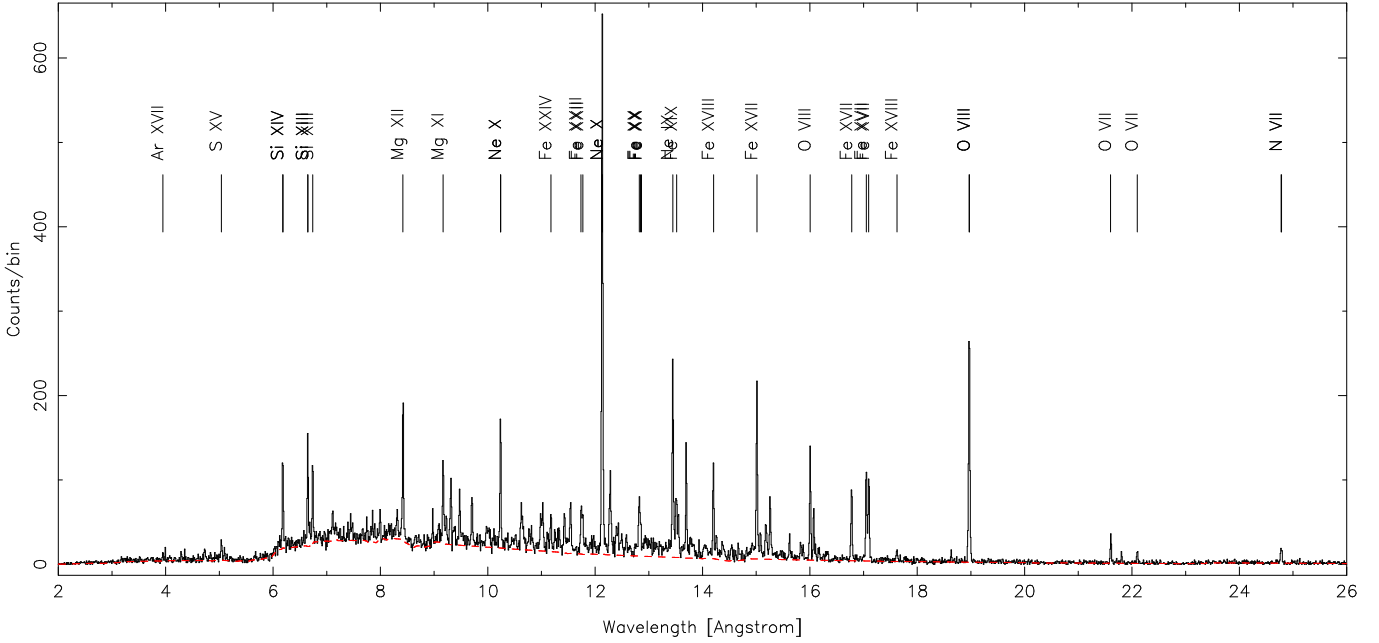}}
 \caption{Chandra/MEG spectrum of AB~Dor. The dashed line represents
 the continuum predicted by the EMD.  
   \label{chandraspec}}
\end{figure*}

AB Dor was observed on 9 October 1999 for 52~ks with the Chandra
High Energy Transmission Grating Spectrograph,
\citep[HETG,][]{wei02}.
The HETGS is made of two
gratings, HEG (High Energy Grating, $\lambda\lambda\sim$1.5--15,
$\lambda/\Delta\lambda\sim$120--1200), and
MEG (Medium Energy Grating $\lambda\lambda\sim$3--30,
$\lambda/\Delta\lambda\sim$60--1200), that operate simultaneously,
permitting the further analysis of the data with different spectral 
resolutions. 
Standard reduction tasks present in the CIAO v2.3 package 
have been employed in the reduction of data retrieved from the Chandra
archive, and the extraction of the HEG and MEG spectra
(Fig.~\ref{chandraspec}).
Two sources are visible in the CCD image at their zero-order positions. 
The main source ($\alpha$=5:28:44.8, $\delta$=--65:26:55.5) is
identified as AB~Dor, while the second source
($\alpha$=5:28:44.4 $\delta$=--65:26:46.5) agrees with the position of
AB~Dor~B (dM4e).  A light curve of AB~Dor A+B was obtained using 
the first orders of HEG and MEG (Fig.~\ref{lightcurves}c), while the
zeroth-order was employed 
to get a light curve of the secondary source alone (the zeroth-order of the
primary source is severely affected by pile-up). No significant
flaring events are present 
in the light curve of AB~Dor B, and a
low-resolution ACIS-S spectrum (sensitivity range 0.4--10.0 keV,
$E/\Delta E\sim$50 at 6~keV) was obtained for it
\citep[see][for further details]{abdor2}. 
This spectrum was employed to calculate the
flux in the range 6--25~\AA\
($f_{\rm X}\sim$1.26$\times$10$^{-12}$~erg~s$^{-1}$~cm$^{-2}$,
$L_{\rm X}\sim$3.35$\times$10$^{28}$~erg~s$^{-1}$). The flux of AB~Dor A+B
calculated in the same spectral range of the MEG spectra is  
$f_{\rm X}\sim$2.75$\times$10$^{-11}$~erg~s$^{-1}$~cm$^{-2}$,
$L_{\rm X}\sim$7.35$\times$10$^{30}$~erg~s$^{-1}$, therefore the secondary
source only represents $\sim$4\% of the total flux of the system. 
A 3-temperature global fit was made to the low-resolution spectrum of
AB~Dor~B \citep{abdor2} using the Astrophysical Plasma Emission Database
  \citep[APED v1.3,][]{aped} in the {\em Interactive Spectral
  Interpretation System} \citep[ISIS,][]{isis} software 
package, provided by the MIT/CXC, and a synthetic MEG
spectrum was 
constructed based on this fit. The comparison of this synthetic
spectrum with the total MEG spectrum shows that the effects of AB~Dor~B
on the total spectrum are negligible (both for HETG 
and XMM/RGS spectra). 

\begin{table}
\caption{XMM observations of AB Dor}\label{times}
\begin{center}
\begin{footnotesize}
\begin{tabular}{lcccrr}
\hline \hline
{Rev.} & {Date} & {pn} & {mos} & {RGS} & {RGS 2}\\
& & & & {$t_{\rm exp}$(ks)}  & cts/s\\
\hline
072 & 1 May 2000 & X & -- & 40+12 & 1.14\\
091 & 7 Jun 2000 & X & -- & 54 & 1.01\\
162 & 27 Oct 2000 & X & X & 57 & 1.22 \\
185 & 11 Dec 2000 & X & X & 56+8+20 & 1.37 \\
205 & 20 Jan 2001 & X & X & 51 & 1.77 \\
266 & 22 May 2001 & X & X & 48 & 1.24 \\
338 & 10 Oct 2001 & -- & X & 38 & 1.21 \\
375 & 26 Dec 2001 & -- & X & 4 & 1.28 \\
429 & 12 Apr 2002 & X & X & 43 & 1.07 \\
462 & 18 Jun 2002 & X & X & 35 & 1.51 \\
532 &  5 Nov 2002 & -- & -- &  9 & 1.60 \\
537 & 15 Nov 2002 & -- & X & 15 & 1.59 \\
546 &  3 Dec 2002 & X & X &  4 & 1.60 \\
560 & 30 Dec 2002 & -- & X & 49 & 1.57 \\
572 & 23 Jan 2003 & -- & -- & 51 & 1.15 \\
\hline
\end{tabular}
\end{footnotesize}
\end{center}
\end{table}

Finally, the emission level of AB~Dor at the time 
of the Chandra observation of AB~Dor has been compared 
with the RGS 2 count rates by folding the Emission Measure
Distribution based on the
Chandra spectra (see below) with the RGS instrumental
response. This simulation yields a count rate of $\sim$1.01 cts/s,
consistent with the count rate obtained in the RGS \#091 observation. 
Further analysis of the light curves and the long-term variability of
the Chandra and XMM observations of AB~Dor will follow in 
\citet{abdor2}.

\section{Data Analysis}\label{sec:analysis}

Stellar coronae are commonly studied through the calculation of the
coronal thermal structure. Such a structure is derived by using the plasma
Emission Measure Distribution vs. Temperature (EMD), with the volume
Emission Measure $EM$($T$) defined as $\int_{\Delta T} N_H N_e
{\rm d}V$ [cm$^{-3}$]. The $EM$ quantifies how much material is
emitting in a temperature range $\Delta T$, and it can be used to 
compute the radiative losses in corona and hence to get information on
the required coronal heating.
Two different approaches are commonly employed in the derivation of the
EMD in the corona: line-based methods and global-fitting techniques. 
The latter are based on the fit
of the whole (lines plus continuum) spectrum using an atomic model and
a discrete number of values 
of temperature and $EM$.  Such an approach uses the metallicity (or even the
abundances of individual elements) as free parameters in the fit.
An alternative method, that can be carried out only for
high-resolution spectra, implies the measurement of individual line
fluxes and its comparison with the fluxes predicted by an atomic model
for a given EMD (``line-based'' methods, or ``EMD reconstruction''). 
The application of these two approaches, although never directly
compared, seems to yield different results, especially regarding the
element abundances \citep{fav03}. 
In this work we will apply a line-based method to reconstruct the
EMD of the corona of AB~Dor in intervals of 0.1 dex in
temperature. We have performed also a global fit to the spectra, with
the aim to compare the results derived with the two techniques.

\subsection{EMD reconstruction}\label{sec:emdmethod}
\begin{figure*}
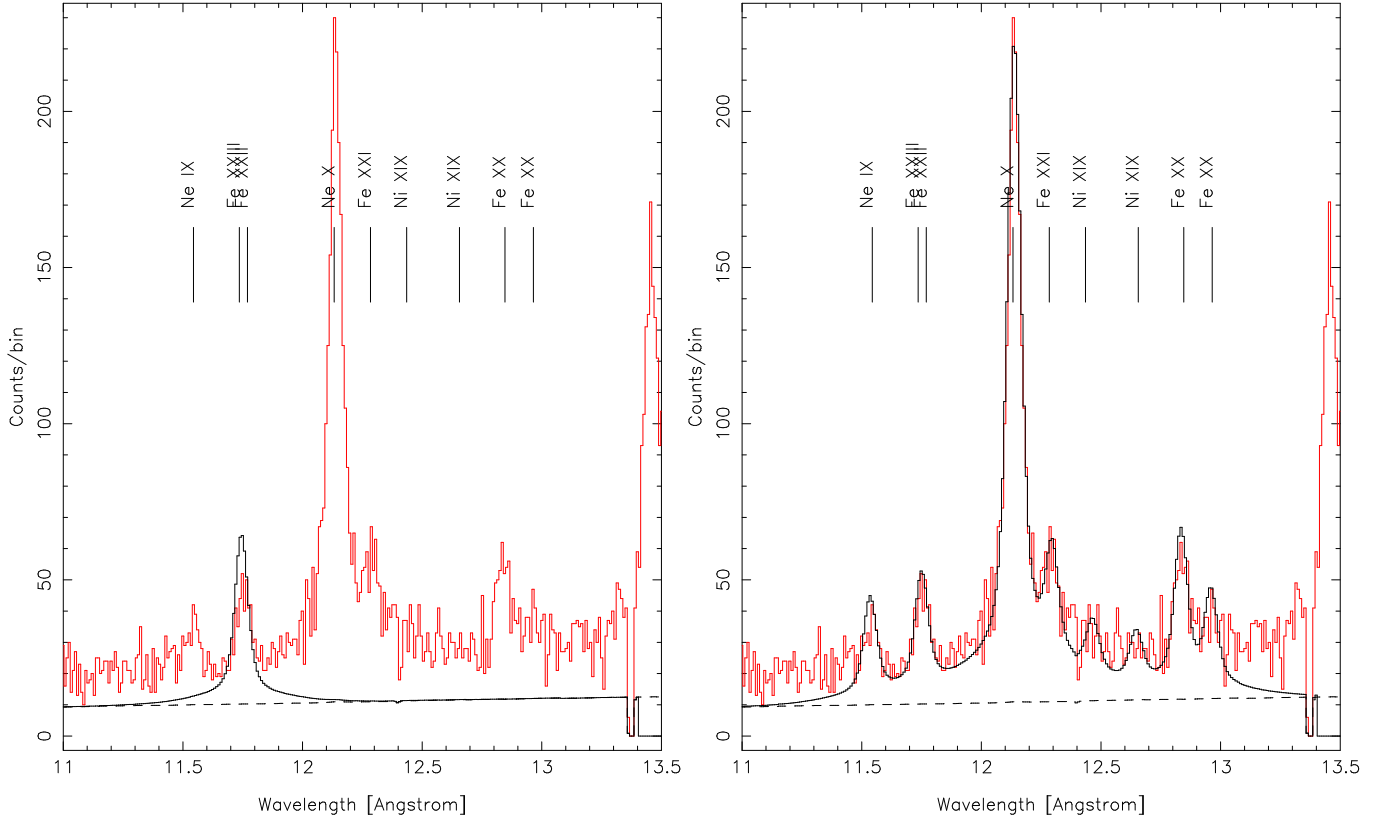

 \resizebox{\hsize}{!}{\includegraphics[angle=270]{H4467F5a.ps}
   \includegraphics[angle=270]{H4467F5b.ps}}
 \caption{Section of the RGS 2 spectrum of rev. \# 091, showing the
   fit to the line \ion{Fe}{xxiii}~$\lambda$11.736 alone (left), and
   including important lines in the same wavelength region
   (right). The flux measured for the \ion{Fe}{xxiii}~$\lambda$11.736
   line is a 43\% higher if the additional lines
   considered in
   the right panel are not included in the fit. 
   The presence of a false continuum created by the LSF of
   adjacent lines influences the measurement of every line. A dashed
   line indicates the continuum predicted from the EMD. 
 \label{lsf}}
\end{figure*}

The EMD reconstruction has been carried out by measuring the fluxes
from spectral
lines present in the RGS and HETG spectra.
RGS spectra are characterized by a Line Spread Function
(LSF) with particularly extended wings that, if not properly taken into
account, may result in a wrong measurement of
the lines fluxes (see an example in Fig.~\ref{lsf}).
The extended wings of the
lines also create a false continuum that makes difficult the placement
of the real source continuum to be used when line fluxes are 
measured (this is especially problematic in the 9--18~\AA\ range,
where numerous lines overlap, see Fig.~\ref{xmmspec}). Such a problem
has been solved by using an iterative process as explained below. The
same process \citep[partly similar to the method described by][]{huen01}
has been applied to the HETG spectral analysis, where the
problems related to the shape of the LSF are less severe. 
Line fluxes from 108 lines, with their corresponding line blends
(Table~\ref{chandrafluxes}), 
have been considered in the analysis of HETG spectra, while 59 lines were
used for the RGS spectra as listed in Table~\ref{xmmfluxes}.
Spectra, response matrices and 
the effective
areas of the instruments, were loaded into the ISIS software 
package, in order to measure the line fluxes, 
following the procedure here described: 

-- A two-temperatures fit to the continuum is made in the case of HETG,
   using line-free regions only, as described in \citet{huen01} and
   \citet{bri01}. This
   fit yields an initial estimate of the continuum, needed for the
   line measurements. In the case of RGS spectra, where line-free
   regions are more difficult to measure, a global 2-T fit to the
   spectrum is performed to calculate
   the initial continuum level.

-- Measurements of line fluxes are made using the continuum predicted
  with the former fit, and convolving delta functions with the Line
  Response Function of the instrument. Simultaneous fit of the MEG and
  HEG spectra were carried out 
  when possible, while separated measurements were made for RGS 1 and
  RGS 2 spectra. 
  Initial line
  identification with atomic data from APED v1.3,  
  is made on the basis of the Emission Measure
  Distribution (EMD) derived by \citet{sanz02} from EUVE data. 
  The false continuum created by the LSF of numerous lines in the
  range 9--18~\AA\ of RGS (see
  Fig.~\ref{lsf}), makes more difficult the measurement of
  lines in this spectral range; a fit
  involving the most intense adjacent lines for each of the line
  measurements was necessary in order to obtain results of the EMD
  that were 
  consistent with the spectra. Much care has to be considered in the
  measurement of any line in this spectral range of RGS. 

-- Predicted fluxes are calculated using the emissivity functions
  present in APED and a trial EMD (the EUVE based EMD mentioned
  above is employed initially in this case), following the method
  described in 
  \citet{dup93}. The quality of the fit is tested with the parameter
  $\beta$\footnote{The parameter $\beta$ is prefered to the
  more standard $\chi^2$ statistic in order to give the same weight to
  all the selected lines; in fact, we note that the uncertainties on
  the comparison between observed and predicted fluxes are dominated
  by systematic errors on the line emissivities rather than by
  poissonian errors on the line counts, and the strongest lines do not
  necessarily have better known atomic data).},  defined as:  
  \begin{equation}\label{eq:beta}
      \beta=\frac{1}{n} \sum_{i=1}^n \Big(\log
      \frac{F_{\mathrm{obs},i}}{F_{\mathrm{pred},i}}\Big)^2
  \end{equation}
  where n is the number of lines considered, $F_{\mathrm{obs}}$ are the 
  fluxes measured in the spectra and $F_{\mathrm{pred}}$ are the
  fluxes predicted by combining the assumed EMD and the APED atomic
  models (a ``perfect'' fit would yield $\beta$=0). 
  Only Fe 
  lines are initially employed, thus avoiding uncertainties due to relative
  abundances. Line blends 
  that may affect the main lines must be included in the analysis. 
  Predicted fluxes are then compared to observed fluxes in order to
  improve the EMD. 

\begin{table*}
\caption{Chandra/HETG line fluxes of AB Dor$^a$}\label{chandrafluxes}
\tabcolsep 3.pt
\begin{scriptsize}
\begin{tabular}{lrrcrrrl}
\hline \hline
 Ion & {$\lambda$$_{\mathrm {model}}$} & {$\lambda$$_{\mathrm {obs}}$} & 
 log $T_{\mathrm {max}}$ & $F_{\mathrm {obs}}$ & S/N & ratio & Blends \\ 
\hline
\ion{Ar}{xvii} &  3.9491 &   3.949 & 7.3 & 4.62e-14 &   4.6 & -0.00 &  \\
\ion{S }{xvi} &  3.9908 &   3.988 & 7.4 & 3.59e-14 &   4.1 &  0.05 & \ion{S}{xvi}  3.9920, \ion{Ar}{xvii}  3.9941, 3.9942, \ion{S}{xv}  3.9980 \\
\ion{S }{xv} &  4.2990 &   4.290 & 7.2 & 3.95e-14 &   4.3 &  0.29 &  \\
\ion{S }{xvi} &  4.7274 &   4.732 & 7.4 & 5.92e-14 &   5.3 & -0.06 & \ion{S }{xvi}  4.7328 \\
\ion{S }{xv} &  5.0387 &   5.039 & 7.2 & 1.05e-13 &   7.1 & -0.13 &  \\
\ion{S }{xv} &  5.1015 &   5.100 & 7.2 & 4.90e-14 &   4.8 & -0.12 & \ion{S }{xv}  5.0983, 5.1025 \\
\ion{Si}{xiv} &  5.2168 &   5.213 & 7.2 & 2.09e-14 &   3.0 &  0.03 & \ion{Si}{xiv}  5.2180 \\
\ion{Si}{xiv} &  6.1804 &   6.183 & 7.2 & 1.47e-13 &  16.4 &  0.06 & \ion{Si}{xiv}  6.1804,  6.1858 \\
\ion{Mg}{xii} &  6.5800 &   6.530 & 7.0 & 1.36e-14 &   5.4 &  0.38 & \ion{Fe}{xxiv}  6.5772, \ion{Mg}{xii}  6.5802 \\
\ion{Si}{xiii} &  6.6479 &   6.648 & 7.0 & 1.42e-13 &  17.3 & -0.06 &  \\
\ion{Si}{xiii} &  6.6882 &   6.687 & 7.0 & 2.62e-14 &   7.5 & -0.07 & \ion{Si}{xiii}  6.6850 \\
No id. &  6.7120 &   6.715 & \ldots & 1.55e-14 &   5.8 & \ldots &  \\
\ion{Si}{xiii} &  6.7403 &   6.741 & 7.0 & 9.77e-14 &  15.7 &  0.07 & \ion{Mg}{xii}  6.7378, \ion{Si}{xiii}  6.7432 \\
\ion{Mg}{xii} &  7.1058 &   7.106 & 7.0 & 2.79e-14 &   8.8 & -0.05 & \ion{Mg}{xii}  7.1069 \\
\ion{Al}{xiii} &  7.1710 &   7.168 & 7.1 & 1.97e-14 &   7.4 &  0.00 & \ion{Fe}{xxiv}  7.1690, \ion{Al}{xiii}  7.1764 \\
\ion{Mg}{xi} &  7.4730 &   7.478 & 6.8 & 1.10e-14 &   5.6 &  0.23 &  \\
\ion{Mg}{xi} &  7.8503 &   7.858 & 6.8 & 2.60e-14 &   8.5 &  0.14 &  \\
\ion{Fe}{xxiv} &  7.9857 &   7.987 & 7.3 & 2.22e-14 &   8.4 &  0.20 & \ion{Fe}{xxiv}  7.9960 \\
\ion{Fe}{xxiii} &  8.3038 &   8.309 & 7.2 & 2.36e-14 &   9.6 &  0.23 &  \\
\ion{Fe}{xxiv} &  8.3761 &   8.377 & 7.3 & 9.81e-15 &   3.2 &  0.33 &  \\
\ion{Mg}{xii} &  8.4192 &   8.416 & 7.0 & 1.32e-13 &  20.5 & -0.18 & \ion{Mg}{xii}  8.4246 \\
\ion{Fe}{xxii} &  8.9748 &   8.974 & 7.1 & 2.30e-14 &   7.8 &  0.15 &  \\
\ion{Ni}{xxvi} &  9.0603 &   9.075 & 7.4 & 1.39e-14 &   6.3 & -0.02 & \ion{Fe}{xxii}  9.0614, \ion{Fe}{xx}  9.0647, 9.0659,  9.0683 \\
No id. &  9.1300 &   9.135 & \ldots & 9.27e-15 &   5.1 & \ldots &  \\
\ion{Mg}{xi} &  9.1687 &   9.170 & 6.8 & 8.85e-14 &  15.8 & -0.16 &  \\
\ion{Fe}{xxi} &  9.1944 &   9.193 & 7.1 & 1.85e-14 &   7.2 &  0.26 & \ion{Mg}{xi}  9.1927, 9.1938, \ion{Fe}{xx}  9.1979 \\
No id. &  9.2190 &   9.216 & \ldots & 1.25e-14 &   5.9 & \ldots & \ion{Ne}{x} 9.215  \\
\ion{Mg}{xi} &  9.2312 &   9.234 & 6.8 & 2.19e-14 &   7.8 & -0.02 & \ion{Mg}{xi}  9.2282 \\
\ion{Ni}{xix} &  9.2540 &   9.251 & 6.9 & 8.05e-15 &   4.7 &  0.24 &  \\
No id. &  9.2870 &   9.291 & \ldots & 1.47e-14 &   6.4 & \ldots &  \ion{Ne}{x} 9.291 \\
\ion{Mg}{xi} &  9.3143 &   9.315 & 6.8 & 4.74e-14 &  11.5 & -0.03 & \\
No id. &  9.3600 &   9.357 & \ldots & 1.24e-14 &   5.9 & \ldots &  \\
\ion{Ni}{xx} &  9.3850 &   9.389 & 7.0 & 1.93e-14 &   7.3 &  0.28 & \ion{Ni}{xxvi}  9.3853, \ion{Ni}{xxv}  9.3900, \ion{Fe}{xxii}  9.3933 \\
\ion{Fe}{xxi} &  9.4797 &   9.473 & 7.0 & 3.95e-14 &  10.4 & -0.16 & \ion{Ne}{x}  9.4807, 9.4809 \\
\ion{Ne}{x} &  9.7080 &   9.705 & 6.8 & 6.20e-14 &  12.8 & -0.14 & \ion{Ne}{x}  9.7085 \\
\ion{Ni}{xix} &  9.9770 &   9.973 & 6.9 & 4.46e-14 &  10.6 & -0.12 & \ion{Ni}{xxv}  9.9700, \ion{Fe}{xxi}  9.9887, \ion{Fe}{xx}  9.9977, 10.0004, 10.0054 \\
No id. & 10.0200 &  10.025 & \ldots  & 1.93e-14 &   6.9 & \ldots &  \\
\ion{Fe}{xx} & 10.1203 &  10.113 & 7.0 & 2.04e-14 &   7.1 & -0.02 & \ion{Ni}{xix} 10.1100, \ion{Fe}{xix} 10.1195, \ion{Fe}{xvii} 10.1210 \\
\ion{Ne}{x} & 10.2385 &  10.238 & 6.8 & 1.55e-13 &  19.3 & -0.17 & \ion{Ne}{x} 10.2396 \\
\ion{Fe}{xxiv} & 10.6190 &  10.625 & 7.3 & 5.65e-14 &  11.2 & -0.01 & \ion{Fe}{xix} 10.6295 \\
\ion{Fe}{xix} & 10.6491 &  10.645 & 6.9 & 2.67e-14 &   7.7 &  0.07 & \ion{Fe}{xix} 10.6414 \\
\ion{Fe}{xxiv} & 10.6630 &  10.665 & 7.3 & 2.70e-14 &   7.7 & -0.12 & \ion{Fe}{xvii} 10.6570 \\
\ion{Fe}{xvii} & 10.7700 &  10.765 & 6.8 & 2.34e-14 &   7.2 & -0.05 & \ion{Ne}{ix} 10.7650, \ion{Fe}{xix} 10.7650 \\
\ion{Fe}{xix} & 10.8160 &  10.818 & 6.9 & 2.46e-14 &   7.3 &  0.06 & \\
\ion{Fe}{xxiii} & 10.9810 &  10.985 & 7.2 & 4.32e-14 &   9.5 & -0.13 &  \\
\ion{Ne}{ix} & 11.0010 &  11.005 & 6.6 & 3.06e-14 &   8.0 &  0.12 & \ion{Ni}{xxii} 10.9920, 10.9927 \\
\ion{Fe}{xxiv} & 11.0290 &  11.035 & 7.3 & 6.73e-14 &  11.8 & -0.08 & \ion{Fe}{xxiii} 11.0190, \ion{Fe}{xvii} 11.0260 \\
\ion{Fe}{xvii} & 11.1310 &  11.128 & 6.8 & 1.84e-14 &   6.1 & -0.09 & \ion{Fe}{xxii} 11.1376 \\
\ion{Fe}{xxiv} & 11.1760 &  11.175 & 7.3 & 6.28e-14 &  11.2 & -0.05 & \ion{Fe}{xxiv} 11.1870 \\
\ion{Fe}{xvii} & 11.2540 &  11.252 & 6.8 & 3.25e-14 &   8.0 & -0.14 & \ion{Fe}{xxiv} 11.2680 \\
\ion{Ni}{xxii} & 11.3049 &  11.303 & 7.1 & 2.37e-14 &   6.8 &  0.17 & \ion{Fe}{xviii} 11.2930, \ion{Fe}{xix} 11.2980, \ion{Ni}{xxi} 11.3180 \\
\ion{Fe}{xviii} & 11.3260 &  11.325 & 6.9 & 3.18e-14 &   7.8 & -0.07 & \ion{Fe}{xxiii} 11.3360 \\
\ion{Fe}{xviii} & 11.4230 &  11.423 & 6.9 & 4.94e-14 &   9.5 &  0.02 & \ion{Fe}{xxii} 11.4270 \\
\ion{Fe}{xxiv} & 11.4320 &  11.435 & 7.3 & 3.39e-14 &   7.7 &  0.01 &  \\
\ion{Fe}{xxii} & 11.4900 &  11.495 & 7.1 & 2.52e-14 &   6.6 &  0.03 & \\
\ion{Fe}{xviii} & 11.5270 &  11.527 & 6.9 & 3.83e-14 &   8.2 &  0.09 & \\
\ion{Ne}{ix} & 11.5440 &  11.543 & 6.6 & 6.28e-14 &  10.5 &  0.06 &  \\
\ion{Fe}{xxiii} & 11.7360 &  11.740 & 7.2 & 7.67e-14 &  11.4 & -0.19 &  \\
\ion{Fe}{xxii} & 11.7700 &  11.773 & 7.1 & 7.12e-14 &  10.8 & -0.22 &  \\
\ion{Fe}{xxii} & 11.8020 &  11.796 & 7.1 & 1.62e-14 &   5.1 &  0.14 & \\
\ion{Ni}{xx} & 11.8320 &  11.818 & 7.0 & 1.80e-14 &   5.5 &  0.04 &  \\
\ion{Fe}{xxii} & 11.9320 &  11.937 & 7.1 & 2.92e-14 &   7.0 &  0.16 &  \\
\ion{Fe}{xxii} & 11.9770 &  11.973 & 7.1 & 2.87e-14 &   6.9 & -0.07 & \ion{Fe}{xxi} 11.9750 \\
\ion{Ne}{x} & 12.1321 &  12.133 & 6.8 & 9.43e-13 &  39.2 & -0.19 & \ion{Fe}{xvii} 12.1240, \ion{Ne}{x} 12.1375 \\
\ion{Fe}{xxiii} & 12.1610 &  12.158 & 7.2 & 7.83e-14 &  11.3 &  0.08 &  \\
\ion{Fe}{xxii} & 12.2100 &  12.210 & 7.1 & 1.99e-14 &   5.6 &  0.02 &  \\
\ion{Fe}{xvii} & 12.2660 &  12.268 & 6.8 & 6.61e-14 &  10.2 &  0.01 &  \\
\ion{Fe}{xxi} & 12.2840 &  12.288 & 7.0 & 1.44e-13 &  15.1 & -0.02 &  \\
\ion{Fe}{xxi} & 12.3930 &  12.398 & 7.0 & 4.99e-14 &   8.8 &  0.23 & \ion{Fe}{xxi} 12.3940 \\
\ion{Fe}{xxi} & 12.4220 &  12.429 & 7.0 & 4.53e-14 &   8.3 & -0.43 & \ion{Fe}{xxii} 12.4311, 12.4318, \ion{Ni}{xix} 12.4350 \\
\ion{Fe}{xxi} & 12.4990 &  12.496 & 7.0 & 3.70e-14 &   7.5 &  0.14 & \ion{Fe}{xxi} 12.4956 \\
No id. & 12.8100 &  12.813 & \ldots  & 6.70e-14 &   9.7 & \ldots &  \\
\ion{Fe}{xx} & 12.8240 &  12.829 & 7.0 & 1.55e-13 &  14.7 & -0.31 & \ion{Fe}{xxi} 12.8220, \ion{Fe}{xx} 12.8460, \ion{Fe}{xx} 12.8640 \\
\ion{Fe}{xx} & 12.9650 &  12.943 & 7.0 & 4.09e-14 &   7.5 & -0.13 & \ion{Fe}{xxii} 12.9530 \\
\ion{Ne}{ix} & 13.4473 &  13.448 & 6.6 & 4.94e-13 &  24.5 &  0.10 & \ion{Fe}{xix} 13.4620 \\
\ion{Fe}{xxi} & 13.5070 &  13.508 & 7.0 & 1.15e-13 &  11.7 &  0.11 & \ion{Fe}{xix} 13.4970 \\
\ion{Fe}{xix} & 13.5180 &  13.518 & 6.9 & 9.93e-14 &  10.9 & -0.11 &  \\
\ion{Fe}{xx} & 13.5350 &  13.535 & 7.0 & 3.03e-14 &   6.0 &  0.19 & \\
\ion{Ne}{ix} & 13.5531 &  13.550 & 6.6 & 9.67e-14 &  10.7 &  0.18 & \ion{Fe}{xix} 13.5510, 13.5540, \ion{Fe}{xx} 13.5583 \\
\ion{Ne}{ix} & 13.6990 &  13.700 & 6.6 & 2.61e-13 &  16.9 &  0.17 & \ion{Fe}{xix} 13.7315, and small blends amounting a 15\% of total flux.  \\
\ion{Ni}{xix} & 13.7790 &  13.768 & 6.8 & 3.91e-14 &   6.5 & -0.18 & \ion{Fe}{xx} 13.7670 \\
\ion{Fe}{xix} & 13.7950 &  13.795 & 6.9 & 5.29e-14 &   7.6 & -0.04 & \\
\hline
\end{tabular}
\end{scriptsize}
\end{table*}
\setcounter{table}{1}
\begin{table*}
\caption{(cont). Chandra/HETG line fluxes of AB Dor$^a$}
\tabcolsep 3.pt
\begin{scriptsize}
\begin{tabular}{lrrcrrrl}
\hline \hline
 Ion & {$\lambda$$_{\mathrm {model}}$} & {$\lambda$$_{\mathrm {obs}}$} & 
 log $T_{\mathrm {max}}$ & $F_{\mathrm {obs}}$ & S/N & ratio & Blends \\ 
\hline
\ion{Fe}{xvii} & 13.8250 &  13.815 & 6.8 & 2.19e-14 &   4.9 & -0.32 &  \\
\ion{Fe}{xix} & 13.8390 &  13.835 & 6.9 & 3.92e-14 &   6.5 &  0.21 & \ion{Fe}{xx} 13.8430 \\
\ion{Fe}{xviii} & 13.9530 &  13.954 & 6.9 & 1.37e-14 &   3.8 & -0.42 & \ion{Fe}{xix} 13.9549, 13.9551, \ion{Fe}{xx} 13.9620 \\
\ion{Fe}{xxi} & 14.0080 &  14.015 & 7.0 & 3.20e-14 &   5.7 &  0.12 &  \\
\ion{Ni}{xix} & 14.0430 &  14.045 & 6.8 & 3.27e-14 &   5.9 & -0.08 &  \\
\ion{Ni}{xix} & 14.0770 &  14.073 & 6.8 & 3.25e-14 &   5.9 & -0.04 &  \\
\ion{Fe}{xviii} & 14.2080 &  14.205 & 6.9 & 2.38e-13 &  15.6 & -0.06 & \\
\ion{Fe}{xviii} & 14.2560 &  14.257 & 6.9 & 7.12e-14 &   8.5 & -0.10 & \ion{Fe}{xx} 14.2670 \\
\ion{Fe}{xviii} & 14.3430 &  14.345 & 6.9 & 3.20e-14 &   5.0 & -0.09 & \\
\ion{Fe}{xviii} & 14.3730 &  14.372 & 6.9 & 6.62e-14 &   7.2 & -0.02 &  \\
\ion{Fe}{xviii} & 14.5340 &  14.535 & 6.9 & 5.64e-14 &   6.5 &  0.04 &  \\
\ion{Fe}{xix} & 14.6640 &  14.675 & 6.9 & 5.34e-14 &   6.3 &  0.17 &  \\
\ion{O }{viii} & 14.8205 &  14.815 & 6.5 & 2.64e-14 &   5.0 & -0.06 &  \\
\ion{Fe}{xx} & 14.9703 &  14.970 & 7.0 & 3.11e-14 &   5.9 &  0.02 & \ion{Fe}{xix} 14.9610 \\
\ion{Fe}{xvii} & 15.0140 &  15.014 & 6.7 & 4.37e-13 &  21.9 & -0.08 &  \\
\ion{Fe}{xx} & 15.0470 &  15.044 & 7.0 & 2.85e-14 &   5.6 &  0.37 & \\
\ion{Fe}{xix} & 15.0790 &  15.078 & 6.9 & 6.72e-14 &   8.5 &  0.21 &  \\
\ion{O }{viii} & 15.1760 &  15.173 & 6.5 & 6.23e-14 &   8.2 & -0.23 & \ion{O }{viii} 15.1765 \\
\ion{Fe}{xix} & 15.1980 &  15.190 & 6.9 & 9.50e-14 &  10.1 &  0.44 &  \\
\ion{Fe}{xvii} & 15.2610 &  15.259 & 6.7 & 2.04e-13 &  14.6 &  0.13 &  \\
\ion{Fe}{xvii} & 15.4530 &  15.455 & 6.7 & 2.39e-14 &   4.9 &  0.16 &  \\
\ion{Fe}{xviii} & 15.6250 &  15.620 & 6.8 & 6.68e-14 &   8.1 &  0.00 &  \\
\ion{Fe}{xviii} & 15.8240 &  15.831 & 6.8 & 5.53e-14 &   7.2 &  0.14 &  \\
\ion{Fe}{xviii} & 15.8700 &  15.860 & 6.8 & 2.49e-14 &   4.9 &  0.03 &  \\
\ion{O }{viii} & 16.0055 &  16.005 & 6.5 & 3.36e-13 &  17.6 & -0.06 & \ion{Fe}{xviii} 16.0040, \ion{O }{viii} 16.0067 \\
\ion{Fe}{xviii} & 16.0710 &  16.071 & 6.8 & 1.63e-13 &  12.2 &  0.27 &  \\
\ion{Fe}{xix} & 16.1100 &  16.105 & 6.9 & 3.66e-14 &   5.8 & -0.14 & \ion{Fe}{xviii} 16.1127 \\
\ion{Fe}{xviii} & 16.1590 &  16.158 & 6.8 & 5.07e-14 &   6.8 &  0.14 &  \\
\ion{Fe}{xvii} & 16.7800 &  16.775 & 6.7 & 2.80e-13 &  15.2 &  0.10 &  \\
\ion{Fe}{xvii} & 17.0510 &  17.053 & 6.7 & 3.67e-13 &  17.0 &  0.19 &  \\
\ion{Fe}{xvii} & 17.0960 &  17.095 & 6.7 & 3.46e-13 &  16.4 &  0.26 &  \\
\ion{Fe}{xviii} & 17.6230 &  17.620 & 6.8 & 5.35e-14 &   6.1 & -0.03 &  \\
\ion{O }{vii} & 18.6270 &  18.635 & 6.3 & 4.16e-14 &   5.0 &  0.12 & \ion{Ar}{xvi} 18.6240 \\
\ion{Ca}{xviii} & 18.6910 &  18.694 & 6.9 & 1.19e-14 &   2.6 & -0.30 &  \\
\ion{O }{viii} & 18.9671 &  18.967 & 6.5 & 1.28e-12 &  26.6 & -0.11 & \ion{O }{viii} 18.9725 \\
\ion{O }{vii} & 21.6015 &  21.608 & 6.3 & 2.23e-13 &   8.6 &  0.09 &  \\
\ion{O }{vii} & 21.8036 &  21.805 & 6.3 & 7.92e-14 &   4.9 &  0.17 &  \\
\ion{O }{vii} & 22.0977 &  22.095 & 6.3 & 1.39e-13 &   6.1 &  0.21 & \ion{Ca}{xvii} 22.1140 \\
\ion{N }{vii} & 24.7792 &  24.779 & 6.3 & 1.31e-13 &   7.2 &  0.00 & \ion{N }{vii} 24.7846 \\
\hline
\end{tabular}

{$^a$ Line fluxes (in erg cm$^{-2}$ s$^{-1}$) of spectral lines
  measured in Chandra/HETG spectra. Measured wavelengths (in \AA) are
  accurate to $\sim10^{-3}$ \AA. log $T_{\mathrm {max}}$ indicates the maximum
  temperature (K) of formation of the line (unweighted by the
  EMD). ``Ratio'' is the log($F_{\mathrm {obs}}$/$F_{\mathrm {pred}}$)  
  of the line. 
  Blends amounting to more than 5\% of the total flux on each line are
  indicated.}
\end{scriptsize}
\end{table*}

\begin{table*}
\caption{XMM line fluxes of AB Dor$^a$}\label{xmmfluxes}
\tabcolsep 3.pt
\begin{scriptsize}
\begin{tabular}{lrccrrcrrl}
\hline \hline
 & & & \multicolumn{3}{c}{XMM rev. 091} & \multicolumn{3}{c}{XMM rev. 205} & \\
 Line ID & {$\lambda$$_{\mathrm {model}}$} & log T$_{\mathrm {max}}$ &
 $F_{\mathrm {obs}}$ & S/N & ratio & $F_{\mathrm {obs}}$ & S/N & ratio
 & Blends \\ 
\hline
\ion{Si}{xiv} &  6.1804 & 7.2 & 1.38e-13 &   3.0 &  0.25 & 2.37e-13 &   5.5 &  0.08 & \\
\ion{Si}{xiii} &  6.6480 & 7.0 & 3.17e-13 &   6.1 &  0.01 & 5.35e-13 &   8.2 & -0.09 & \ion{Si}{xiii}  6.6882, 6.7403 \\
\ion{Mg}{xii} &  8.4192 & 7.0 & 1.52e-13 &   7.3 & -0.04 & 4.25e-13 &  13.0 &  0.00 & \\
\ion{Mg}{xi} &  9.1687 & 6.8 & 9.60e-14 &  13.8 & -0.13 & 2.10e-13 &  19.4 & -0.13 &  \\
\ion{Mg}{xi} &  9.2312 & 6.8 & 1.56e-13 &   8.1 &  0.20 & 3.06e-13 &  10.5 &  0.17 & \ion{Ni}{xix}  9.2540, \ion{Mg}{xi}  9.3143\\
\ion{Ne}{x} &  9.7080 & 6.8 & 1.06e-13 &   7.6 & -0.08 & 2.44e-13 &  11.4 &  0.02 & \\
\ion{Ne}{x} & 10.2385 & 6.8 & 2.28e-13 &  13.8 & -0.15 & 4.41e-13 &  18.1 & -0.12 & \\
\ion{Fe}{xxiv} & 10.6190 & 7.3 & 1.41e-13 &   5.8 &  0.11 & 2.71e-13 &   7.4 &  0.07 &  \ion{Fe}{xix} 10.6491, 10.6840, \ion{Fe}{xxiv} 10.6630 \\
\ion{Fe}{xvii} & 10.7700 & 6.8 & 8.34e-14 &   5.9 &  0.12 & 1.76e-13 &   5.3 &  0.22 & \ion{Fe}{xix} 10.8160 \\
\ion{Fe}{xxiii} & 11.0190 & 7.2 & 1.50e-13 &  11.1 &  0.03 & 3.46e-13 &  11.6 &  0.03 & \ion{Fe}{xxiii} 10.9810, \ion{Ne}{ix} 11.0010, \ion{Fe}{xxiv} 11.0290 \\
\ion{Fe}{xxiv} & 11.1760 & 7.3 & 1.21e-13 &   9.8 &  0.19 & 1.66e-13 &   6.8 & -0.02 & \ion{Fe}{xvii} 11.1310\\
\ion{Fe}{xvii} & 11.2540 & 6.8& \ldots & \ldots & \ldots & 1.60e-13 &   6.7 &  0.23 & \ion{Fe}{xxiii} 11.2850 \\
\ion{Fe}{xviii} & 11.4230 & 6.9 & 9.49e-14 &   7.9 &  0.04 & 2.55e-13 &   9.5 &  0.13 & \ion{Fe}{xxiv} 11.4320, \ion{Fe}{xviii} 11.4494, \ion{Fe}{xxiii} 11.4580\\
\ion{Fe}{xviii} & 11.5270 & 6.9 & 1.55e-13 &  11.3 &  0.00 & 2.11e-13 &   8.2 & -0.11 & \ion{Fe}{xxii} 11.4900, \ion{Ni}{xix} 11.5390, \ion{Ne}{ix} 11.5440\\
\ion{Fe}{xxiii} & 11.7360 & 7.2 & 2.35e-13 &  15.5 &  0.10 & 5.09e-13 &  16.1 &  0.01 & \ion{Fe}{xxii} 11.7700 \\
\ion{Ne}{x} & 12.1321 & 6.8 & 1.40e-12 &  50.3 &  0.02 & 2.34e-12 &  52.9 & -0.22 & \\
\ion{Fe}{xxi} & 12.2840 & 7.0 & 2.33e-13 &  14.4 &  0.05 & 5.20e-13 &  12.7 &  0.07 & \ion{Fe}{xvii} 12.2660 \\
\ion{Ni}{xix} & 12.4350 & 6.9 & 1.23e-13 &  10.0 & -0.19 & 2.09e-13 &   8.0 & -0.13 & \ion{Fe}{xxi} 12.3930, 12.4220\\
\ion{Fe}{xxi} & 12.4990 & 7.0& \ldots & \ldots & \ldots & 1.87e-13 &   7.8 &  0.11 & \ion{Fe}{xx} 12.5260, 12.5760 \\
\ion{Fe}{xxi} & 12.6490 & 7.0 & 7.78e-14 &   7.4 &  0.17 & 1.45e-13 &   6.7 &  0.30 & \ion{Fe}{xx} 12.6210, \ion{Fe}{xxii} 12.6307, \ion{Ni}{xix} 12.6560 \\
\ion{Fe}{xx} & 12.8240 & 7.0 & 2.83e-13 &  18.9 & -0.05 & 5.73e-13 &  18.1 & -0.04 & \\
\ion{Fe}{xx} & 12.9650 & 7.0 & 1.62e-13 &  13.0 &  0.01 & 3.52e-13 &  12.5 &  0.06 & \ion{Fe}{xx} 12.9120, 12.9920, 13.0240, \ion{Fe}{xix} 12.9330, 13.0220 \\
\ion{Ne}{ix} & 13.4473 & 6.6 & 7.72e-13 &  40.6 &  0.10 & 1.11e-12 &  45.7 &  0.02 & \ion{Fe}{xx} 13.3850\\
\ion{Fe}{xix} & 13.5180 & 6.9 & 3.00e-13 &  22.2 & -0.08 & 7.01e-13 &  36.7 &  0.04 & \ion{Fe}{xxi} 13.5070, \ion{Ne}{ix} 13.5531 \\
\ion{Ne}{ix} & 13.6990 & 6.6 & 4.25e-13 &  16.2 &  0.14 & 7.04e-13 &  20.1 &  0.15 & \ion{Fe}{xix} 13.6450, 13.7315, 13.7458 \\
\ion{Fe}{xix} & 13.7950 & 6.9 & 1.95e-13 &  14.2 & -0.00 & 1.60e-13 &   9.0 & -0.28 & \ion{Ni}{xix} 13.7790, \ion{Fe}{xvii} 13.8250 \\
\ion{Fe}{xx} & 13.9620 & 7.0 & 7.23e-14 &   6.8 &  0.10 & 8.48e-14 &   7.8 & -0.07 & \ion{Fe}{xviii} 13.9530, \ion{Fe}{xix} 13.9525, 13.9546, 13.9571, 13.9743 \\
\ion{Fe}{xxi} & 14.0080 & 7.0 & 1.62e-13 &   8.7 & -0.00 & 1.69e-13 &  10.9 & -0.11 & \ion{Fe}{xix} 14.0340, 14.0717, \ion{Ni}{xix} 14.0430, 14.0770 \\
\ion{Fe}{xviii} & 14.2080 & 6.9 & 4.03e-13 &  21.2 & -0.04 & 6.34e-13 &  26.4 & -0.06 & \ion{Fe}{xviii} 14.2560 \\
\ion{Fe}{xviii} & 14.3730 & 6.9 & 1.72e-13 &  14.1 & -0.11 & 2.57e-13 &  19.7 & -0.16 & \ion{Fe}{xx} 14.3318, 14.4207, \ion{Fe}{xviii} 14.3430, 14.4250, 14.4392 \\
\ion{Fe}{xviii} & 14.5340 & 6.9 & 1.89e-13 &  12.5 &  0.17 & 2.91e-13 &  18.4 &  0.15 &  \ion{Fe}{xviii} 14.4856, 14.5056, 14.5710, 14.6011 \\
\ion{O}{viii} & 14.8205 & 6.5 & 8.62e-14 &   8.9 &  0.06 & 1.45e-13 &  10.3 &  0.00 & \ion{Fe}{xviii} 14.7820, \ion{O}{viii} 14.8207, \ion{Fe}{xx} 14.8276 \\
\ion{Fe}{xvii} & 15.0140 & 6.7 & 6.08e-13 &  32.8 & -0.10 & 9.49e-13 &  37.5 & -0.14 & \ion{Fe}{xix} 15.0790 \\
\ion{O}{viii} & 15.1760 & 6.5 & 1.91e-13 &  15.1 &  0.04 & 3.17e-13 &  16.9 & -0.02 & \ion{Fe}{xix} 15.1980 \\
\ion{Fe}{xvii} & 15.2610 & 6.7 & 2.02e-13 &  18.7 &  0.05 & 3.59e-13 &  22.6 &  0.07 &  \\
\ion{Fe}{xviii} & 15.6250 & 6.8 & 8.21e-14 &   9.5 &  0.01 & 1.57e-13 &  12.5 &  0.08 &  \\
\ion{Fe}{xviii} & 15.8700 & 6.8 & 1.20e-13 &  10.7 &  0.20 & 1.83e-13 &  13.3 &  0.17 & \ion{Fe}{xviii} 15.8240 \\
\ion{O}{viii} & 16.0066 & 6.5 & 3.71e-13 &  29.9 & -0.12 & 6.71e-13 &  32.6 & -0.14 & \ion{O}{viii} 16.0055 \\
\ion{Fe}{xviii} & 16.0710 & 6.8 & 1.88e-13 &  15.9 & -0.06 & 2.89e-13 &  16.2 & -0.09 & \ion{Fe}{xviii} 16.0450, 16.1590, \ion{Fe}{xix} 16.1100, 16.1590 \\
\ion{Fe}{xvii} & 16.7800 & 6.7 & 2.93e-13 &  22.4 &  0.03 & 4.58e-13 &  26.6 & -0.00 &  \\
\ion{Fe}{xvii} & 17.0510 & 6.7 & 7.11e-13 &  36.9 &  0.13 & 1.03e-12 &  50.1 &  0.07 & \ion{Fe}{xvii} 17.0960 \\
\ion{Fe}{xviii} & 17.6230 & 6.8 & 6.80e-14 &   9.1 & -0.01 & 1.31e-13 &  12.4 &  0.07 &  \\
\ion{O}{vii} & 18.6270 & 6.3 & 3.91e-14 &   5.7 & -0.25 & 1.19e-13 &  10.5 & -0.18 & \ion{Ca}{xviii} 18.6910 \\
\ion{O}{viii} & 18.9671 & 6.5 & 1.33e-12 &  54.4 & -0.20 & 2.91e-12 &  79.2 & -0.16 & \\
\ion{N}{vii} & 20.9095 & 6.3 & 4.06e-14 &   4.7 &  0.05 & 5.13e-14 &   4.6 & -0.02 & \ion{N}{vii} 20.9106 \\
\ion{Ca}{xvi} & 21.4500 & 6.7 & 2.02e-14 &   7.2 &  0.11 & 7.14e-14 &   6.2 &  0.14 & \ion{Ca}{xvi} 21.4410 \\
\ion{O}{vii} & 21.6015 & 6.3 & 2.99e-13 &  27.7 &  0.08 & 5.24e-13 &  22.0 &  0.08 & \ion{Ca}{xvi} 21.6100 \\
\ion{O}{vii} & 21.8036 & 6.3 & 9.69e-14 &   8.7 &  0.18 & 1.60e-13 &  10.7 &  0.16 & \ion{Ca}{xvii} 21.8220 \\
\ion{O}{vii} & 22.0977 & 6.3 & 1.77e-13 &  12.9 &  0.17 & 3.38e-13 &  16.8 &  0.19 & \ion{Ca}{xvii} 22.1140 \\
\ion{Ar}{xvi} & 23.5460 & 6.7 & 3.57e-14 &   2.9 &  0.16 & 4.92e-14 &   3.6 & -0.15 & \ion{Ar}{xvi} 23.5900, \ion{Ca}{xvi} 23.6260 \\
\ion{N}{vii} & 24.7792 & 6.3 & 2.18e-13 &  21.8 & -0.04 & 3.80e-13 &  28.9 &  0.03 & \ion{Ar}{xvi} 24.8540 \\
\ion{Ar}{xvi} & 24.9910 & 6.7 & 1.71e-14 &   2.6 & -0.15 & 6.44e-14 &  11.5 &  0.00 & \ion{Ar}{xvi} 25.0130, \ion{Ar}{xv} 25.0500 \\
\ion{C}{vi} & 26.9896 & 6.2 & 2.66e-14 &   5.7 &  0.29 & 4.44e-14 &   6.9 &  0.18 & \ion{C}{vi} 26.9901 \\
\ion{C}{vi} & 28.4652 & 6.2 & 2.48e-14 &   5.3 & -0.20 & 7.48e-14 &   9.8 & -0.05 & \ion{C}{vi} 28.4663 \\
\ion{S}{xiv} & 30.4270 & 6.5 & 2.93e-14 &   5.6 & -0.18 & 4.77e-14 &   6.9 & -0.14 & \ion{S}{xiv} 30.4690, \ion{Ca}{xi} 30.4710 \\
\ion{S}{xiv} & 32.4160 & 6.5 & 2.13e-14 &   3.1 &  0.15 & 3.07e-14 &   3.4 &  0.20 &  \\
\ion{S}{xiv} & 32.5600 & 6.5 & 3.93e-14 &   4.8 &  0.12 & 4.69e-14 &   4.7 &  0.09 & \ion{S}{xiv} 32.5750 \\
\ion{C}{vi} & 33.7342 & 6.1 & 1.86e-13 &  17.8 & -0.08 & 3.54e-13 &  24.1 & -0.12 & \ion{C}{vi} 33.7396 \\
\hline
\end{tabular}

$^a$  Line fluxes (in erg cm$^{-2}$ s$^{-1}$) of spectral lines
  measured in XMM/RGS first order spectra. log $T_{\mathrm {max}}$
  indicates the maximum 
  temperature (K) of formation of the line (unweighted by the
  EMD). ``Ratio'' is the log($F_{\mathrm obs}$/$F_{\mathrm {pred}}$)
  of the line.
  Blends amounting to more than 5\% of the total flux for each line are
  indicated (see also blends in Table~\ref{chandrafluxes}).
\end{scriptsize}
\end{table*}

-- Once the $\beta$ parameter converges to a minimum value, neon
  lines are added to 
  the analysis in order to extend the EMD to lower temperatures. The
  \ion{Ne}{x} lines are mostly formed in a temperature range 
  (log~$T$[K]$\sim$6.6--7.3)
  which 
  overlaps with that of the Fe lines, therefore permitting to set
  the Ne abundance. Then, the oxygen abundance can be set employing the
  \ion{O}{viii} lines, and the \ion{O}{vii} lines provide information
  down to log~$T$(K)$\sim$6.2.
  Finally, the rest of the elements (Mg, S, Si, Ar, Ni, N, Ca, C, Al) are
  added one by one in the analysis, in 
  order to calculate the abundances (relative to Fe) that better fit
  their fluxes, leaving the EMD unchanged. 

-- The continuum is recomputed, and new flux measurements are
  performed. Changes of the EMD shape are now permitted, once all
  elements are included 
  in the fit.  An iterative process is followed until the measurements
  of the lines converge. Electron
  densities (see below) are included in the
  analysis by applying the relevant values in their
  corresponding temperature ranges (log~$T$[K]$\sim$6.1--6.4 for density
  derived from \ion{O}{vii} lines, log~$T$[K]$\sim$6.5--6.6 for density
  from \ion{Ne}{ix} lines, and log~$T$[K]$\sim$6.7--7.2 for density from
  \ion{Mg}{xi}, \ion{Fe}{xxi} and \ion{Fe}{xxii} line ratios). The Fe
  abundance is determined once the EMD has been calculated, and the
  rest of abundances are scaled to the [Fe/H] value.

Statistical uncertainties on the EMD values (Table~\ref{tab:emds}),
are estimated using a 
Montecarlo method that varies the line fluxes randomly by 1--$\sigma$
(1000 different sets of fluxes are tested),
and calculates the best result among 1000 possible EMDs (randomly
generated, including variations by up to 1--$\sigma$ over the calculated
abundances) for each pattern of fluxes. A criterion of convergence
is established in the improvement of $\beta$ by at least 5$\times10^{-4}$. 
The 68\% of the central values among those resulting from this fit,
are considered in order to set the 1--$\sigma$ error
bars of the $EM$ values. These error bars are not
independent,  a higher value of the $EM$ in a given temperature bin usually
requires a lower value in an adjacent bin in order to match the
observed line fluxes. 
Finally, uncertainties in the determination of the abundances
(Table~\ref{tab:abundances})
are evaluated
considering not only the statistical errors of the measured fluxes,
but also the dispersion observed in the 
$F_{\mathrm{obs}}/F_{\mathrm{pred}}$ ratio of
all the lines of each element. This is an indirect way to evaluate the
errors induced by the uncertainties in atomic models. 

\begin{figure}
 \resizebox{\hsize}{!}{\includegraphics{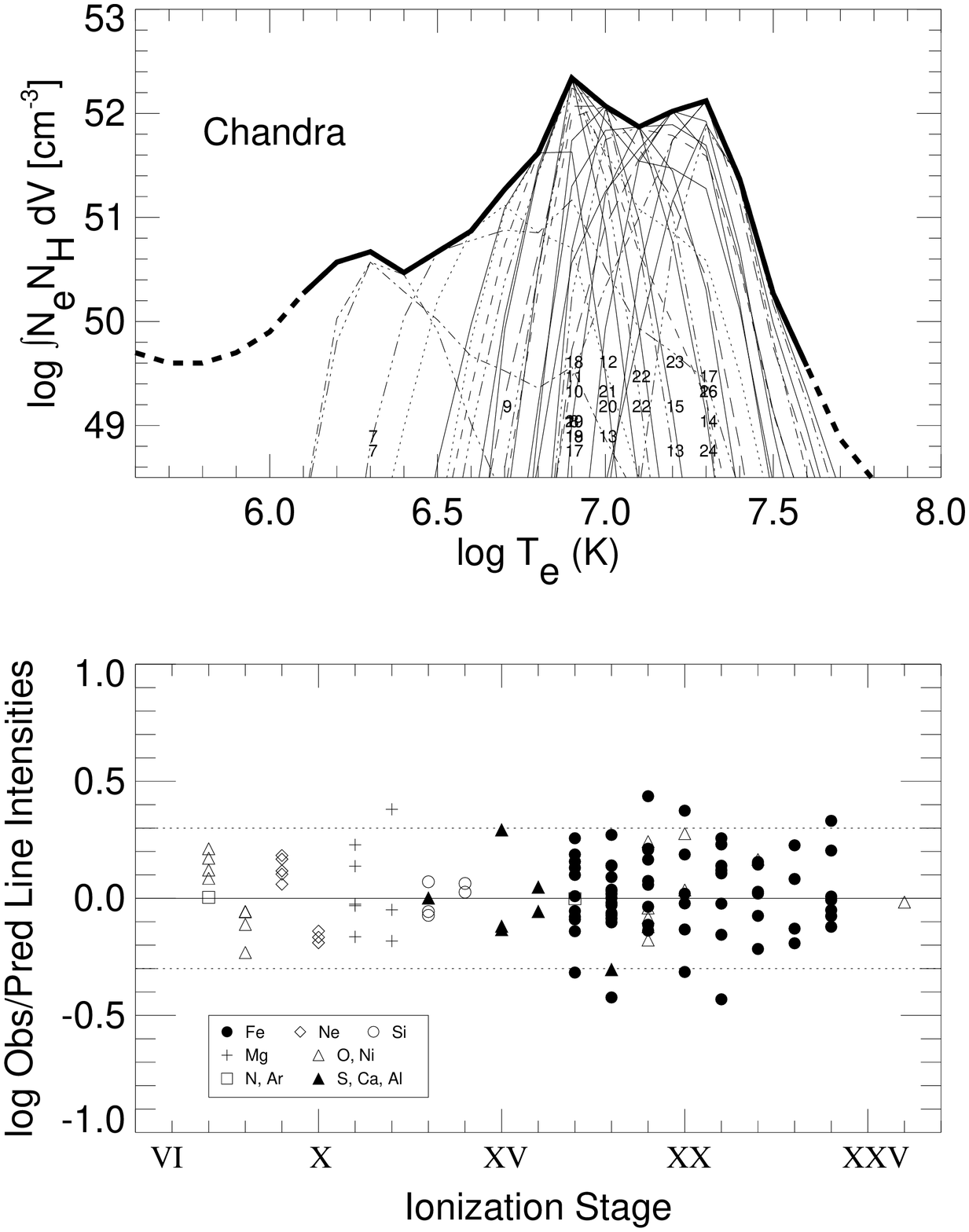}}
 \caption{{\em Upper}: EMD derived from Chandra/HETG data. Thin lines
   represent the relative contribution function for each ion (the
   emissivity function multiplied by the EMD at each point). Small
   numbers indicate the ionization stages of the species. {\it Lower}:
   Observed to predicted line flux ratios for the ion stages in the upper
   figure. The dotted lines denote a factor of 2.\label{chandraemd}}
\end{figure}

\begin{figure}
 \resizebox{\hsize}{!}{\includegraphics{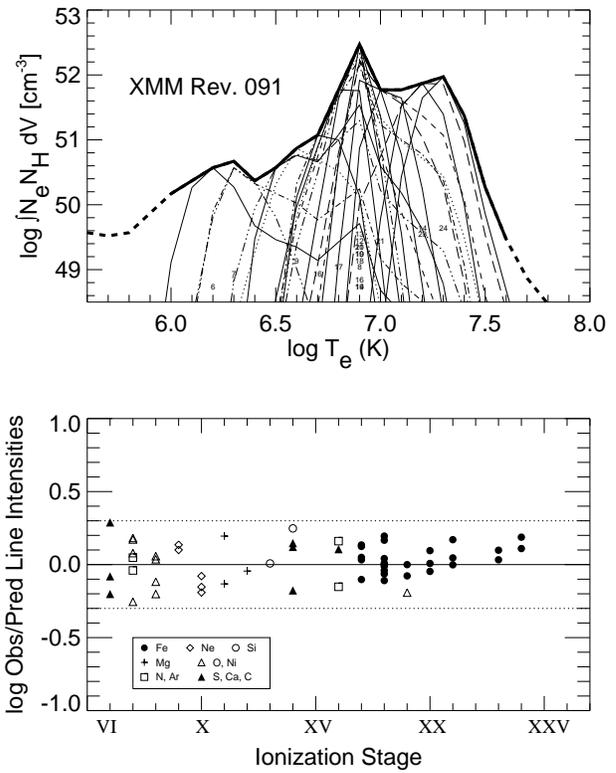}}
 \caption{Same as Fig.~\ref{chandraemd}, for the XMM/RGS data during
   rev. \# 091.\label{xmmemd}}
\end{figure}

The results
obtained in the RGS rev. \#091 are consistent with those of the
HETG observation.  The bump present at log~$T$(K)$\sim$6.9 is very
robust in both cases (see Table~\ref{tab:emds}). 
The presence of a hot tail is well
established for log~$T$(K)$\sim$7.0--7.3, with some uncertainties on the
exact shape.  Larger error bars in the EMD are present at
log~$T$(K)$\la$6.6, due to the lack of Fe 
lines that would reduce the uncertainties in the abundances of Ne and O.
\ion{Fe}{xv} and \ion{Fe}{xvi} lines formed at those temperatures are well 
observed with EUVE \citep{sanz02} for AB~Dor, and can be used (with
some caution) for a consistency test of the results.
However, these lines are affected by
uncertainties in the determination of the ISM absorption, and the EUVE
spectrum could correspond to a different level of emission. 
The formal solution, shown in Figs.~\ref{chandraemd} and \ref{xmmemd} and
Table~\ref{tab:emds},
has been a compromise of the results found for the Ne lines and the
mentioned EUVE lines, that are overestimated by up to $\sim$50\%
with the solution derived from HETG spectra.  
Finally, abundances of elements like Ca, Al and Ar are not very robust 
since they are derived from little number of lines. Also, C and N
lines present in the 
spectra have a temperature range that overlaps mostly with that of
\ion{O}{vii}, and hence the abundances of C and N are linked to that of
O. Marginal
inconsistency in the abundances calculated from RGS and HETG detectors
is only found for Ca, N and Ne. 

\begin{figure}
 \resizebox{\hsize}{!}{\includegraphics{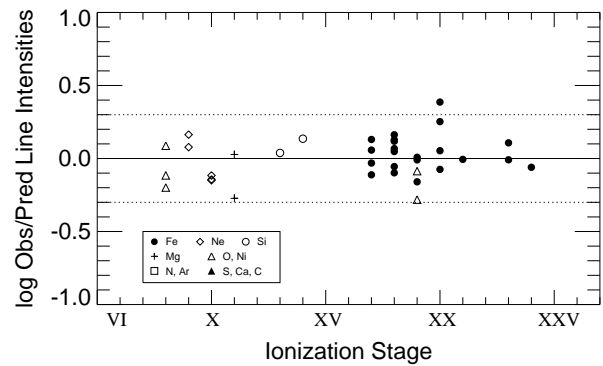}}
 \caption{Flux ratios corresponding to the measurements of the second
   order of the XMM/RGS data during rev.~\#091 (see
   Table~\ref{tab:2ndorder}), using the EMD and abundances derived
   from the line fluxes in the first order.\label{emdsecond}}
\end{figure}

The measurements of line fluxes in the RGS spectra during rev.~\#205
yield an EMD with similar shape to that of HETG and RGS rev.\#091, but
with higher $EM$ values. Abundances of the elements did not change
significantly between 
the two RGS observations, except for the worse constrained cases
of Ca and N. 
Finally, the line fluxes have been measured in the second order of RGS
during rev.~\#091 (Table~\ref{tab:2ndorder}),
resulting in a very good agreement with the EMD and abundances calculated
with the first order of RGS (Fig.~\ref{emdsecond}). 

\begin{table}
\caption{Emission Measure Distribution of AB Dor}\label{tab:emds}
\begin{center}
\begin{tabular}{lccc}
\hline \hline
log $T$ & \multicolumn{3}{c}{log $\int N_{\rm e} N_{\rm H} {\rm d}V$ (cm$^{-3}$)$^a$} \\
(K) & HETG & RGS 091 & RGS 205 \\
\hline
6.1 & 50.27: & 50.37: & 50.47: \\
6.2 & 50.57$^{+0.30}_{-0.20}$  & 50.57$^{+0.20}_{-0.20}$  & 50.67$^{+0.15}_{-0.35}$  \\
6.3 & 50.67$^{+0.20}_{-0.20}$  & 50.67$^{+0.15}_{-0.25}$  & 50.77$^{+0.10}_{-0.30}$  \\
6.4 & 50.47$^{+0.20}_{-0.30}$  & 50.37$^{+0.30}_{-0.40}$  & 50.52$^{+0.30}_{-0.30}$  \\
6.5 & 50.67$^{+0.20}_{-0.40}$  & 50.57$^{+0.40}_{-0.30}$  & 50.77$^{+0.40}_{-0.30}$  \\
6.6 & 50.87$^{+0.20}_{-0.40}$  & 50.87$^{+0.40}_{-0.20}$  & 51.07$^{+0.30}_{-0.40}$  \\
6.7 & 51.27$^{+0.10}_{-0.30}$  & 51.07$^{+0.20}_{-0.30}$  & 51.42$^{+0.20}_{-0.30}$  \\
6.8 & 51.62$^{+0.10}_{-0.20}$  & 51.77$^{+0.00}_{-0.00}$  & 52.17$^{+0.05}_{-0.05}$  \\
6.9 & 52.34$^{+0.03}_{-0.07}$  & 52.47$^{+0.00}_{-0.00}$  & 52.57$^{+0.00}_{-0.00}$  \\
7.0 & 52.07$^{+0.10}_{-0.10}$  & 51.77$^{+0.05}_{-0.25}$  & 52.32$^{+0.05}_{-0.05}$  \\
7.1 & 51.87$^{+0.20}_{-0.20}$  & 51.77$^{+0.05}_{-0.35}$  & 52.27$^{+0.10}_{-0.10}$  \\
7.2 & 52.02$^{+0.10}_{-0.30}$  & 51.87$^{+0.20}_{-0.30}$  & 52.27$^{+0.10}_{-0.30}$  \\
7.3 & 52.12$^{+0.05}_{-0.15}$  & 51.97$^{+0.20}_{-0.20}$  & 52.37$^{+0.10}_{-0.30}$  \\
7.4 & 51.37$^{+0.40}_{-0.20}$  & 51.37$^{+0.40}_{-0.20}$  & 51.27$^{+0.40}_{-0.40}$  \\
7.5 & 50.27$^{+0.40}_{-0.30}$  & 50.27$^{+0.40}_{-0.30}$  & 50.27$^{+0.40}_{-0.40}$  \\
7.6 & 49.57: & 49.47: & 49.67: \\
\hline
\end{tabular}
\end{center}
$^a$Emission Measure, where $N_{\rm e}$ 
and $N_{\rm H}$ are electron and hydrogen densities, in cm$^{-3}$.
\end{table}

\subsection{Global fitting approach}\label{sec:global}
The EMD 
calculated for the XMM observation in rev. \# 091 can be compared to
the results obtained 
by \citet{gud01} by applying a global fit to the spectrum. These
authors found that the spectrum could be fitted using 3 values of the
$EM$ (log~$EM$(cm$^{-3}$)$\sim$51.92, 52.56, 52.52) at the temperatures
log~$T$(K)=6.57, 6.90 and 7.35 respectively, using the VMEKAL atomic model. 
However, at the time of their analysis only preliminary calibration of
the RGS data was available, and a direct comparison of results is not
possible. 
In order to understand whether the use of a global fit using a 
3-T model affects the results,
we tried a 3-T fit to the RGS rev.~\#~091 spectrum, using APED in
ISIS, as we made for the EMD reconstruction. 
The fit was performed first using only one temperature and
[Fe/H] as free parameters, and adding progressively a second and third
temperature components, and the abundances of individual elements. 
Several sets of values yield a similar quality of the fit, depending on the
initial values and on the element abundances that are permitted to
vary. These results show different values
of abundances and emission measure (sometimes inconsistent among them
as well), although the three temperatures do not vary substantially
($\Delta$log~$T$[K]$\sim\pm$0.1). 
The best fit found was obtained for 
log~$T$(K)=6.52$\pm$0.02, 6.89$\pm$0.01 and 7.27$\pm$0.01, and
log~$EM$(cm$^{-3}$)$\sim$52.23$\pm$0.03, 52.61$\pm$0.02 
and 52.54$\pm$0.01, together with the abundances listed in
Table~\ref{tab:abundances}.
These values were then used to predict the line fluxes to be compared
with the measured ones.
A value of $\beta\sim$0.0479 is obtained if all the spectral lines are
considered, but the result improves to $\beta\sim$0.0161 (a value
which is
close to the $\beta\sim$0.0151 obtained with the EMD reconstruction)
when the \ion{Ca}{xvi} $\lambda$21.45 line is excluded
(this line is off by more than one order of magnitude in the
3-T fit).  
However some abundances obtained with the 3-T fit are not
consistent with those calculated with the line-based approach
(Table~\ref{tab:abundances}).
Moreover, the flux
of the EUVE lines of \ion{Fe}{xv} $\lambda$284 and \ion{Fe}{xvi}
$\lambda$334 and $\lambda$361 lines is overestimated by one order of
magnitude, a result that we consider inconsistent:
variations in the flux of the lines between the different observations
are possible, but in the case of the mentioned \ion{Fe}{xv}  
and \ion{Fe}{xvi} lines, 
no enhancements of more than a factor of 4 were detected
during very large flares in stars with a coronal
structure similar to that of AB~Dor \citep{sanz02}.

The same procedure was applied for a simultaneously fitting of the
Chandra HEG 
and MEG spectra using a 3-T model as above, finding as the best result
log~$T$(K)$\sim$6.60$\pm$0.03, 6.91$\pm$0.01, 7.26$\pm$0.02,
and log~$EM$(cm$^{-3}$)$\sim$52.33$\pm$0.06, 52.62$\pm$0.03,
52.64$\pm$0.02 respectively, and the abundances listed in
Table~\ref{tab:abundances}. 
Moreover, the results are somewhat dependent on whether 
the abundances of elements with few lines in the
spectrum, like Al, are treated as individual free parameters or fixed
to the solar value.  In this case the comparison of the observed fluxes
and those resulting from the 3-T fit shows a much larger dispersion 
($\beta$=0.0614) than for the EMD reconstruction ($\beta$=0.0247).
Chandra results are 
inconsistent with those of the RGS observation in rev. \#~091, unlike
in the line-based approach. 

\begin{table*}
\caption{XMM/RGS second order line fluxes of AB Dor$^a$ (rev.~\#091)}\label{tab:2ndorder}
\tabcolsep 3.pt
\begin{scriptsize}
\begin{tabular}{lrcrrrl}
\hline \hline
 Ion & {$\lambda$$_{\mathrm {model}}$} &  
 log $T_{\mathrm {max}}$ & $F_{\mathrm {obs}}$ & S/N & ratio & Blends \\ 
\hline
\ion{Si}{xiv} &  6.1804 & 7.2 & 1.06E--13 &   4.0 &  0.14 & \\
\ion{Si}{xiii} &  6.6480 & 7.0 & 3.40E--13 &   4.9 &  0.04 & \ion{Si}{xiii}  6.6882, 6.7403 \\
\ion{Mg}{xi} &  9.1687 & 6.8 & 1.39E--13 &   5.6 &  0.03 &  \\
\ion{Mg}{xi} &  9.3143 & 6.8 & 3.14E--14 &   3.6 & -0.27 & \ion{Ni}{xxv}  9.3400 \\
No id. &  9.4790 &  & 9.63E--14 &   6.1 & \ldots &  \\
\ion{Ne}{x} &  9.7080 & 6.8 & 9.17E--14 &   6.2 & -0.14 & \\
\ion{Ne}{x} & 10.2385 & 6.8 & 2.31E--13 &   7.8 & -0.15 & \ion{Ne}{x} 10.2396 \\
\ion{Fe}{xix} & 10.6491 & 6.9 & 7.57E--14 &   3.7 & -0.16 & \ion{Fe}{xxiv} 10.6190, 10.6630, \ion{Fe}{xix} 10.6414, 10.6840, \ion{Fe}{xvii} 10.6570\\
\ion{Fe}{xxiii} & 11.0190 & 7.2 & 1.37E--13 &  10.5 & -0.01 & \ion{Fe}{xxiii} 10.9810, \ion{Ne}{ix} 11.0010, \ion{Fe}{xvii} 11.0260, \ion{Fe}{xxiv} 11.0290 \\
\ion{Fe}{xxiv} & 11.1760 & 7.3 & 6.85E--14 &   3.7 & -0.06 & \ion{Fe}{xvii} 11.1310, \ion{Fe}{xxiv} 11.1870 \\
\ion{Fe}{xviii} & 11.4230 & 6.9 & 1.16E--13 &   5.0 &  0.13 & \ion{Fe}{xxii} 11.4270, \ion{Fe}{xxiv} 11.4320, \ion{Fe}{xviii} 11.4494, \ion{Fe}{xxiii} 11.4580 \\
\ion{Fe}{xviii} & 11.5270 & 6.9 & 1.81E--13 &   6.7 &  0.07 & \ion{Fe}{xxii} 11.4900, \ion{Ni}{xix} 11.5390, \ion{Ne}{ix} 11.5440 \\
\ion{Fe}{xxiii} & 11.7360 & 7.2 & 2.40E--13 &   8.3 &  0.11 & \ion{Fe}{xxii} 11.7700 \\
\ion{Ne}{x} & 12.1320 & 6.8 & 1.66E--12 &  25.4 & -0.12 & \ion{Ne}{x} 12.1375 \\
\ion{Fe}{xxi} & 12.2840 & 7.0 & 2.06E--13 &   9.7 & -0.01 & \ion{Fe}{xvii} 12.2660 \\
\ion{Ni}{xix} & 12.4350 & 6.9 & 1.00E--13 &   4.4 & -0.28 & \ion{Fe}{xxi} 12.3930, 12.4220 \\
\ion{Fe}{xx} & 12.8240 & 7.0 & 2.65E--13 &  14.7 & -0.07 & \ion{Fe}{xxi} 12.8220, \ion{Fe}{xx} 12.8460, 12.8640 \\
\ion{Fe}{xx} & 12.9650 & 7.0 & 1.30E--13 &   5.9 &  0.25 & \ion{Fe}{xix} 12.9311, 12.9330, 12.9450, \ion{Fe}{xxii} 12.9530 \\
\ion{Ne}{ix} & 13.4473 & 6.6 & 7.33E--13 &  26.2 &  0.08 & \ion{Fe}{xix} 13.4620 \\
\ion{Fe}{xix} & 13.5180 & 6.9 & 3.65E--13 &  17.1 &  0.01 & \ion{Fe}{xix} 13.4970, \ion{Fe}{xxi} 13.5070, \ion{Ne}{ix} 13.5531 \\
\ion{Ne}{ix} & 13.6990 & 6.6 & 4.54E--13 &  19.6 &  0.16 & \ion{Fe}{xix} 13.6450, 13.7315, 13.7458 \\
\ion{Fe}{xix} & 13.7950 & 6.9 & 1.68E--13 &   6.4 & -0.01 & \ion{Ni}{xix} 13.7790, \ion{Fe}{xvii} 13.8250 \\
\ion{Fe}{xx} & 13.9620 & 7.0 & 1.41E--13 &   9.7 &  0.39 & \ion{Fe}{xix} 13.9525, 13.9546, 13.9549, 13.9551, 13.9571, 13.9743, \ion{Fe}{xviii} 13.9530 \\
\ion{Ni}{xix} & 14.0770 & 6.8 & 1.10E--13 &   4.3 & -0.09 & \ion{Ni}{xix} 14.0430, \ion{Fe}{xix} 14.0717 \\
\ion{Fe}{xviii} & 14.2080 & 6.9 & 3.02E--13 &  12.1 & -0.06 & \\
\ion{Fe}{xx} & 14.2670 & 7.0 & 1.13E--13 &   4.1 &  0.05 & \ion{Fe}{xviii} 14.2560, \ion{Fe}{xviii} 14.2560 \\
\ion{Fe}{xviii} & 14.3730 & 6.9 & 1.18E--13 &   6.5 & -0.10 & \ion{Fe}{xx} 14.3318, \ion{Fe}{xviii} 14.3430\\
\ion{Fe}{xvii} & 15.0140 & 6.7 & 5.72E--13 &  26.1 & -0.11 & \ion{Fe}{xix} 15.0790 \\
\ion{O}{viii} & 15.1760 & 6.5 & 2.15E--13 &   8.1 &  0.09 & \ion{Fe}{xix} 15.1980 \\
\ion{Fe}{xvii} & 15.2610 & 6.7 & 1.68E--13 &   8.0 & -0.03 &  \\
\ion{Fe}{xviii} & 15.6250 & 6.8 & 1.17E--13 &   8.5 &  0.16 &  \\
\ion{Fe}{xviii} & 15.8700 & 6.8 & 1.01E--13 &   7.3 &  0.12 & \ion{Fe}{xviii} 15.8240 \\
\ion{O}{viii} & 16.0066 & 6.5 & 3.71E--13 &  12.3 & -0.12 & \ion{Fe}{xviii} 16.0040, \ion{O}{viii} 16.0055 \\
\ion{Fe}{xviii} & 16.0710 & 6.8 & 2.43E--13 &   8.5 &  0.05 & \ion{Fe}{xviii} 16.0450, 16.1590, \ion{Fe}{xix} 16.1100 \\
\ion{Fe}{xvii} & 16.7800 & 6.7 & 3.10E--13 &   8.0 &  0.06 &  \\
\ion{Fe}{xvii} & 17.0510 & 6.7 & 7.04E--13 &  16.4 &  0.13 & \ion{Fe}{xvii} 17.0960 \\
\ion{O}{viii} & 18.9671 & 6.5 & 1.34E--12 &  15.0 & -0.20 & \ion{O}{viii} 18.9725 \\
\hline
\end{tabular}

{$^a$ Line fluxes (in erg cm$^{-2}$ s$^{-1}$) 
  measured in XMM/RGS second order spectra. 
  log $T_{\mathrm {max}}$ indicates the maximum
  temperature (K) of formation of the line (unweighted by the
  EMD). ``Ratio'' is the log($F_{\mathrm {obs}}$/$F_{\mathrm {pred}}$) 
  of the line. 
  Blends amounting to more than 5\% of the total flux for each line are
  indicated.}
\end{scriptsize}
\end{table*}

In summary, both the lined-based method and the global fit approaches
permit, only in the case of the RGS 
spectrum, to achieve a solution that is consistent with the observed
spectra. 
However, the error bars provided by the global-fitting technique are 
unrealistic given the wide range of solutions that are ``statistically
acceptable''. The global-fitting technique assumes that the model can
explain {\em all} the lines in the spectrum, while the line-based
approach can reject lines that are problematic and assumes only to
know well some of the lines. 
In the particular case of the 3-T approach, 
the use of 3 temperatures to
characterize a corona can be only a parameterization of the real
multi-temperature coronal structure.
The fit of the Chandra spectrum clearly demonstrates 
that a more detailed thermal
structure is necessary to explain the HETG emission. 
We therefore discourage the use of models with few isothermal
components whenever the measurement of individual lines is
possible. Hereafter we will refer  only to the results obtained using
the line-based approach.

\subsection{Electron density}

\begin{figure}
 \resizebox{\hsize}{!}{\includegraphics[angle=90]{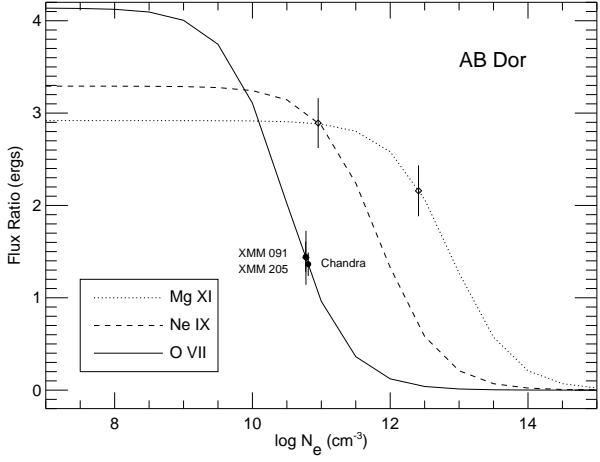}}
 \caption{Electron densities from He-like triplets (see
 Table~\ref{tab:densHe}).\label{denselec}}
\end{figure}

He-like triplets observed in the spectral range covered by HETG and
RGS can be employed in the calculations of the
electron densities \citep[see][and references therein]{ness02}. 
The relevant lines correspond to 
the resonance ($r$), intercombination ($i$), and forbidden ($f$)
transitions. The $f$/$i$ flux ratio can be employed to calculate the
electron density, while the ($f$+$i$)/$r$ flux ratio gives the average
temperature of formation for each triplet. 
The \ion{O}{vii} ($\lambda$21.6015, $\lambda$21.8036,
$\lambda$22.0977), \ion{Ne}{ix} ($\lambda$13.4473, $\lambda$13.5531,
$\lambda$13.6990),  
\ion{Mg}{xi} ($\lambda$9.1687, $\lambda$9.2282+$\lambda$9.2312,
$\lambda$9.3143) and \ion{Si}{xiii}
($\lambda$6.6479, $\lambda$6.6882, $\lambda$6.7403) triplets have been
measured in the  
HETG spectra (Fig.~\ref{denselec}, Table~\ref{tab:densHe}), 
and only the \ion{O}{vii} triplet is used in RGS
due to heavy blending in the other triplets. In all cases there are
some blends that need to be 
considered, and they may eventually affect the results of the analysis (see
Tables~\ref{chandrafluxes}, \ref{xmmfluxes}). 
Some of these blends were measured as separated lines, 
while in other cases they are included in the measurement of the
observed fluxes, and then evaluated using the EMD and abundances
calculated. 
Only in the case of the \ion{Si}{xiii} the $f$/$i$ flux ratio
(3.9$\pm$0.5) resulted 
inconsistent with the predicted values ($\sim$2.9 in the low-density
limit). This might indicate a problem in the
atomic models lacking line blends to the $\lambda$6.7402 line. Given
the uncertainties in the determination of the Si and Mg abundances, a
contamination of up to $\sim$18\% of the flux can be attributed to
\ion{Mg}{xii} lines. A contamination of $\sim$25\% to the
$\lambda$6.7402 line would be necessary in order to 
to have a $f$/$i$ value consistent with the low-density
limit. A slightly higher contamination ($\sim$26\%) would be required
to reach a value consistent with 
a density of log n$_e$(cm$^{-3}$)$\sim$12.5, as obtained from 
the \ion{Fe}{xxi} and \ion{Fe}{xxii} lines (see Sect.~\ref{sec:resemd}).

\begin{figure}
 \resizebox{\hsize}{!}{\includegraphics[angle=90]{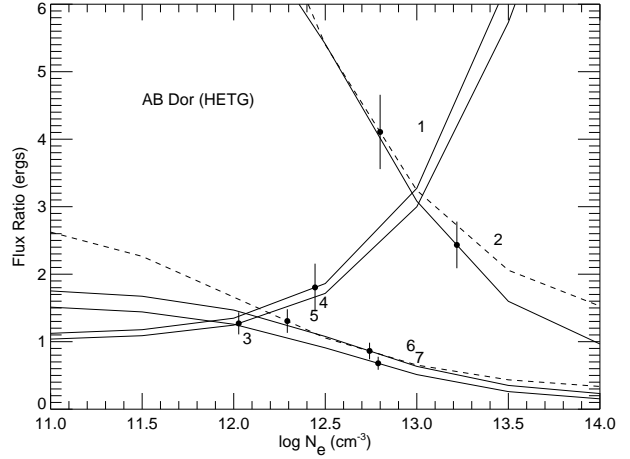}
   }
 \caption{Electron densities from \ion{Fe}{xxi} (dashed lines) and
   \ion{Fe}{xxii} (solid lines)
   lines ratios, as indicated in Table~\ref{tab:densFe}.\label{denselecFe}}
\end{figure}

Electron densities can be calculated also from several flux ratios
of \ion{Fe}{xxi} and \ion{Fe}{xxii} lines (with maximum contribution
at log~$T$[K]$\sim$7.0 and 7.1 respectively) present in the HETG
spectrum (Fig.~\ref{denselecFe}, Table~\ref{tab:densFe}). Lines
selected for these ratios are 
little, or not affected at all, by blends present in the APED models,
and hence we consider the results from these ratios more trustful than
those obtained from the He-like triplets. These results are consistent
with those calculated using \ion{Fe}{xxi} and \ion{Fe}{xxii} lines ratios
in the EUV range \citep{sanz02}.

\section{Results}\label{sec:results}

\subsection{EMD and electron densities}\label{sec:resemd}
\begin{figure}
 \resizebox{\hsize}{!}{\includegraphics[angle=90]{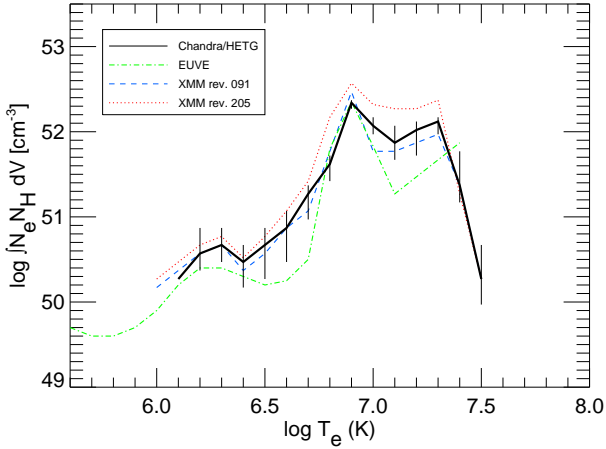}}
 \caption{Comparison of EMDs derived from different instruments. The EMD
 from EUVE observations \citep{sanz02} was scaled for a coronal
 abundance of [Fe/H]=$-0.57$. Representative $1-\sigma$ error bars are
 shown for the Chandra/HETG EMD only (note that these error bars are
 not independent, see text). \label{allemds}} 
\end{figure}

The general shape of the
EMD derived from XMM and Chandra is consistent with the EMD obtained
with EUVE \citep{sanz02}, indicating a
corona dominated by multi-T plasma with a peak at log~$T$(K)$\sim$6.9,
very well constrained by the line
fluxes (see Fig.~\ref{allemds}). The EMD at higher temperatures
is supported by a large number of lines from different elements
identified in the 
HETG and RGS spectra.
Finally a lower peak seems to be present around 
log~$T$(K)$\sim$6.3. However, the lack of coverage of Fe lines makes 
the results for the EMD at log~$T$(K)$\la$6.6 less robust than for
higher temperatures. 

Comparison of the EMDs reconstructed at different times
(Fig.~\ref{allemds}), including those of the XMM observations
corresponding to different levels of activity of AB~Dor, shows that
the EMD peak is very stable, and thus an increase in
the emission level is linked to higher emission measure values at all
temperatures. The EMDs based on Chandra and XMM data show a steep
increase with temperature in the range log~$T$(K)$\sim$6.4--6.9, with a
slope comprised between $T^4$ and $T^5$, while they are almost flat
from the peak up to log~$T$(K)$\sim$7.3.

\begin{table*}
\caption{Abundances of the elements ([X/H], solar units)}\label{tab:abundances}
\tabcolsep 3.pt
\begin{center}
\begin{footnotesize}
\begin{tabular}{lccccrrr}
\hline \hline
{X} & {FIP} & HETG & HETG & XMM 091 & XMM 091 & XMM 205 \\
    &  eV   & (3-T) & (EMD) &  (3-T) & (EMD) & (EMD) \\
\hline
Al &  5.98 & -0.62$\pm$0.33 & -0.35$\pm$0.13 & \ldots & \ldots & \ldots \\
Ca &  6.11 & -0.90$\pm$0.59 &  0.23$\pm$0.63 & -1.07$\pm$0.50 &  0.38$\pm$0.26 & 0.62$\pm$0.32 \\
Ni &  7.63 & -0.87$\pm$0.18 & -0.20$\pm$0.11 & -0.41$\pm$0.10 & -0.08$\pm$0.37 & -0.28$\pm$0.29 \\
Mg &  7.64 & -0.69$\pm$0.04 & -0.39$\pm$0.12 & -0.46$\pm$0.05 & -0.43$\pm$0.17 & -0.34$\pm$0.14 \\
Fe &  7.87 & -0.85$\pm$0.03 & -0.57$\pm$0.04 & -0.67$\pm$0.02 & -0.57$\pm$0.07 & -0.57$\pm$0.07 \\
Si &  8.15 & -0.60$\pm$0.04 & -0.47$\pm$0.09 & -0.40$\pm$0.08 & -0.39$\pm$0.29 & -0.37$\pm$0.17 \\
S  & 10.36 & -0.50$\pm$0.11 & -0.20$\pm$0.16 & -0.57$\pm$0.08 & -0.32$\pm$0.22 & -0.49$\pm$0.21 \\
C  & 11.26 & \ldots         & \ldots         & -0.36$\pm$0.04 & -0.02$\pm$0.23 & -0.05$\pm$0.16 \\
O  & 13.61 & -0.55$\pm$0.04 & -0.15$\pm$0.11 & -0.54$\pm$0.02 & -0.05$\pm$0.11 & -0.03$\pm$0.09 \\
N  & 14.53 & -0.40$\pm$0.22 & -0.04$\pm$0.14 & -0.28$\pm$0.05 &  0.20$\pm$0.13 & 0.10$\pm$0.11 \\
Ar & 15.76 & -0.49$\pm$0.41 &  0.05$\pm$0.22 & -0.42$\pm$0.16 & -0.05$\pm$0.33 & 0.09$\pm$0.20 \\
Ne & 21.56 & -0.21$\pm$0.03 &  0.13$\pm$0.10 & -0.03$\pm$0.02 &  0.28$\pm$0.10 & 0.25$\pm$0.11 \\
\hline
\end{tabular}
\end{footnotesize}
\end{center}
\end{table*}

The electron densities calculated using the \ion{O}{vii} triplet 
(log~$n_{\rm e}$[cm$^{-3}$]$\sim$10.7--10.9 at log~$T$[K]$\sim$6.3)
are consistent in the three observations of HETG and RGS
(Table~\ref{tab:densHe}). The value obtained
for rev.~\#091 is also close to the one
(log~$n_{\rm e}$[cm$^{-3}$]$\sim$10.3$^{+0.3}_{-0.4}$) reported by 
\citet{gud01}.
The density calculated from the
\ion{Ne}{ix} triplet (log~$n_{\rm e}$[cm$^{-3}$]$\sim$11 at
log~$T$[K]$\sim$6.6) is the value most affected 
by the presence of lines not deblended,
and evaluated using the atomic model
combined with the EMD \citep[see also][]{bri01}, and hence it has to
be considered more uncertain. Higher densities
(log~$n_{\rm e}$[cm$^{-3}$]$\ga$12 at log~$T$[K]$\sim$6.9) are indicated by
the \ion{Mg}{xi} triplet.   Finally,
the \ion{Fe}{xxi} and \ion{Fe}{xxii} densities
(Table~\ref{tab:densFe}) are obtained from lines that are
quite well measured, and with little contributions from blends.
Values in the range  log~$n_{\rm e}$(cm$^{-3}$)$\sim$12.3--12.8 are
indicated by most of these lines which form at $T\sim$10~MK, having
excluded the two most discrepant values. 
The results clearly suggest an increase of 
the electron density with temperature, also
observed for Capella \citep{bri01,argi03}, and
formerly suggested by EUVE observations of Capella and other active
stars \citep[see][and references therein]{bri96,drake96,sanz02}.

In particular, the results obtained with the
\ion{Mg}{xi} triplet and the \ion{Fe}{xxi} and \ion{Fe}{xxii} line
ratios seem to confirm the findings already derived from EUVE
\citep{dup93,bri96,sanz02,sanz03} and Chandra \citep{huen01,huen03} data, 
with densities of at least 
log~$n_{\rm e}$(cm$^{-3}$)$\sim$12 at log~$T$(K)$\sim$6.8--7.1.

\subsection{Element abundances}
\begin{figure}
  \resizebox{\hsize}{!}{\includegraphics{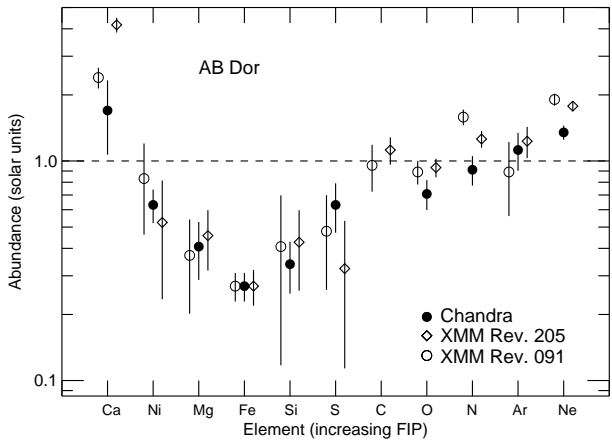}}
  \caption{Element abundances in the corona of AB Dor, with respect to solar
    photosphere. A dashed line indicates the solar photospheric
      abundance \citep{anders}.\label{abundances}}
\end{figure}

Coronal element abundances are subject of some controversy 
because of the different results (in many cases contradictory)
found in the comparison from observations with different instruments
and/or analyzed with different techniques \citep[see][and references
  therein]{fav03}. Stars with low levels of activity tend to show
similar trends with the First Ionization Potential
(FIP) as in the solar case \citep[see][and references
  therein]{bow00}, with elements with low FIP ($<$10 eV; e.g. Mg,
Si, Fe) being enhanced with respect to elements with higher FIP ($\ge$ 10
eV; e.g. O, Ne, Ar) when compared with the photospheric values. 
The opposite effect is observed instead for more active stars \citep[the so
called MAD ---``Metal abundances deficient''---
effect, or the ``inverse FIP effect''][]{sch96,brink01,dra01}. However, many 
of the studied active stars lack accurate information on the
photospheric abundances, and/or show contradictory results depending
on the technique applied in the analysis. We have
carried out the two most typical approaches for the coronal analysis
(global-fitting to the spectrum and EMD reconstruction from line
fluxes) in order to clarify what is the origin of this discrepancy. 
The use of the EMD reconstruction approach shows very consistent
results (see Table~\ref{tab:abundances} and Fig.~\ref{abundances}) for
the different instruments, 
while the global-fitting results are not convergent,
and show contradictory results (e.g. for Ni, Mg, Fe or Ne, see
Table~\ref{tab:abundances}) in HETG and RGS for observations with
similar flux level. 
The line-based analysis applied to Chandra and XMM data show a clear
and consistent pattern of the abundances vs. FIP. An 
initial depletion of the abundances seems to be present
for elements with FIP increasing up to the Fe value, 
and a progressive increment of the abundances for the
elements with higher FIP \citep[a similar behavior seems to be
present in the case of AR Lac,][]{huen03}. 
Photospheric abundances of AB~Dor 
are near solar photospheric values \citep[with {[Fe/H]}$\sim$0.1][]{vil87b}. 
Coronal Fe abundance depletion for AB~Dor was already found from EUVE
and ASCA observations \citep{ruc95,mew96}, now confirmed by the HETG
and RGS analysis. 

The EMD can be extended down to log~$T$(K)$\sim$4 through the
measurements of lines in the UV. \citet{sanz02} reported measurements
from IUE not simultaneous to those of EUVE, and constructed an EMD of
AB~Dor in the range log~$T$(K)$\sim$4.0--7.3. The calculation of the 
EMD in the lower
temperature region (approximately log~$T$[K]$\sim$4.0--5.5) was
dominated basically by C lines, while the determination of the EMD for
log~$T$(K)$\ga$5.6 depends on Fe lines only. The combined data of IUE
and EUV, in absence of better information, indicated a minimum of the
EMD around log~$T$(K)$\sim$5.8. 
The calculation of the [C/Fe] abundance with RGS allows us to correct
the position of this minimum.
Assuming that the value of
[C/Fe]$\sim$0.5 is constant in the whole range of temperature, and the
general flux levels of these observations is similar, the EMD
calculated in the corona by \citep{sanz02} should be 0.5 dex higher,
and the minimum of the EMD should occur at a lower temperature,
log~$T$(K)$\sim$5.3. Measurements of lines formed in the region
log~$T$(K)$\sim$5--6 would be needed in order to nail the temperature of
this minimum. 

\section{Discussion and conclusions}\label{sec:conclusions}

Our analysis of the Chandra/HETG and XMM/RGS spectra has provided
consistent results in terms of the plasma emission measure
distribution (EMD) vs. temperature, the abundances of individual elements in
corona, and even the plasma density as determined from the \ion{O}{vii}
He-like triplet. The superior spectral resolution of the Chandra
instrument has also allowed us to obtain reliable estimates of the plasma 
density at different temperatures using the \ion{Ne}{ix} and \ion{Mg}{xi}
triplets, as well as density-sensitive \ion{Fe}{xxi} and \ion{Fe}{xxii}
line ratios. 
The corona of AB Dor appears to have a quite stable thermal structure 
with an amount of plasma steeply increasing with temperature 
from log~$T$(K)$\sim 6.4$, to 
log~$T$(K)$\sim 6.9$. A substantial amount of plasma (within a factor 2
from the peak value) is also present in a plateau of the EMD extending 
up to $T \sim 2 \times 10^7$\,K.
A less constrained secondary peak 
is possibly also present at $T \sim 2 \times 10^6$\,K, 
the characteristic temperature of the corona of the quiet Sun. 

The above description applies to AB~Dor in quiescent state, i.e. outside of
any evident isolated flaring event, and it is essentially in agreement
with the 
information already available from previous observations of AB~Dor with
EUVE. However, the quiescent X-ray emission is not steady, but
characterized by significant variability on time scales shorter than the
rotation period, suggestive of
a very dynamic corona where a large number of small-scale flares may occur
at any time. On the other hand,
we recall that AB~Dor is also capable of producing 
very strong flares, which may affect the coronal thermal structure quite
significantly: in the extreme case of the flares observed by SAX
\citep{mag00}, the peak value of the total volume emission measure 
was of the order of $10^{54}$\,cm$^{-3}$ at a temperature $T\sim 10^8$\,K.

By comparing the EMDs corresponding to the lowest and highest
X-ray emission levels observed in the first three years of XMM-Newton
observations we learn that
an increase of the source luminosity (in the range 6--20~\AA) 
from $L_{\rm x} = 6.8\times
10^{29}$\,erg s$^{-1}$ 
(Jun 2000) to $L_{\rm x} = 1.2\times 10^{30}$\,erg s$^{-1}$ (Jan
2001), not associated 
to any large flare, can be explained by
an increase of the whole EMD by factors 1.2--3, with the largest variation
occurring in the level of the high-temperature plateau of the EMD. 
In principle, larger emission measures can be obtained with a (linear)
increase  
of the emitting volume, or a (square root) increase of the plasma density, 
or both.  
Unfortunately, the statistical uncertainties on the density derived from 
the analysis of the \ion{O}{vii} triplet at the two epochs does not allow us 
to distinguish between these possibilities.

Plasma densities are a key parameter to try to interpret the above
scenario in terms of classes of magnetically-confined coronal
structures. The measurements of the \ion{O}{vii} triplet
(log~$n_{\rm e}$[cm$^{-3}$]$\sim$10.8) yield plasma
pressures $p_{\rm e} \sim 20-40$ dyn cm$^{-2}$ 
at $T \sim 1$--$2 \times 10^6$\,K, already quite high for solar standards.
Even higher values are indicated by a number of
other line ratio diagnostics derived from the Chandra/HETG data only.
In particular, the \ion{Ne}{ix} triplet (log~$n_{\rm e}$[cm$^{-3}$]$\sim$11)
indicates a pressure $p_{\rm e}
\sim 10^{2}$\,dyn cm$^{-2}$ at $T \sim 3$--$5 \times 10^6$\,K, while
the \ion{Mg}{xi} triplet (log~$n_{\rm e}$[cm$^{-3}$]$\sim$12.4) and the
density-sensitive  
\ion{Fe}{xxi} and \ion{Fe}{xxii} lines
(log~$n_{\rm e}$[cm$^{-3}$]$\sim$12.5) suggest a pressure increasing from 
$p_{\rm e} \sim 2 \times 10^{3}$\,dyn cm$^{-2}$ up to 
$\sim 0.7$--$2 \times 10^{4}$\,dyn cm$^{-2}$ 
for plasma temperatures in the range $5 \times 10^6$--$2 \times 10^7$\,K.

\begin{figure}
  \resizebox{\hsize}{!}{\includegraphics[angle=90]{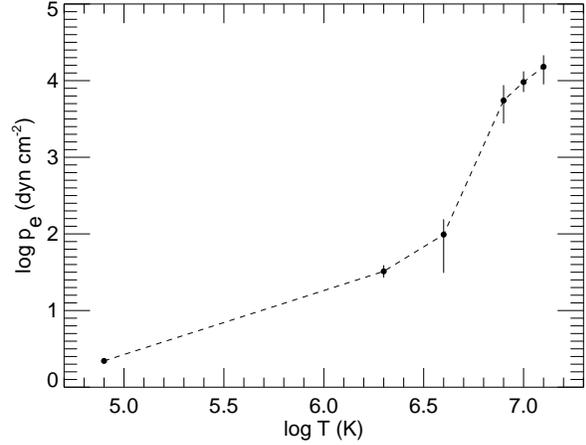}}
  \caption{Electron pressure derived from the electron densities at
    different temperatures (see text). Last two points represent
    averages over different density values in
    Table~\ref{tab:densFe}.\label{pressure}}
\end{figure}

Taking into account also the results obtained by \citet{sch98}
and by \citet{ake00} from \ion{C}{iii} line ratios 
-- log~$n_{\rm e}$(cm$^{-3}$)$\sim$11, yielding $p_{\rm e} \sim 2.2$\,dyn
cm$^{-2}$ at   
$T \sim 8 \times 10^4$\,K, i.e.\ at the base of the transition region --
we derive the picture illustrated in Fig.~\ref{pressure}: the plasma pressure
appears to increase steadily from the transition region to the corona,
with a steeper and steeper piece-wise power-law dependence on
temperature (from $T^{0.7}$ at low temperatures, up to $T^5$ in the
range of the EMD plateau). Note that this relationship was
found using the peak temperatures of the line emissivity
functions weighted by the EMD; a slightly different but qualitatively
consistent behavior would appear considering the effective temperatures 
of line formation provided by the $(f+i)/r$ ratios, available for the 
He-like triplets only.

\begin{table}
\caption{Electron densities calculated from He-like triplets}\label{tab:densHe}
\tabcolsep 3.pt
\begin{center}
\begin{footnotesize}
\begin{tabular}{lccccl}
\hline \hline
{Ion} & {($f+i$)/$r$} & {$T_{\rm ratio}$} & $f$/$i$ & {log $n_{\rm e}$(cm$^{-3}$)} & {Instrument} \\
\hline
\ion{O}{vii} & 0.83$\pm$0.06 & 6.24$\pm$0.07& 1.44$\pm$0.17 & 10.77$\pm0.08$ & RGS (091)\\
\ion{O}{vii} & 0.78$\pm$0.05 & 6.33$\pm$0.06 & 1.36$\pm$0.13 & 10.82$\pm0.06$ & RGS (205)\\
\ion{O}{vii} & 0.86$\pm$0.15 & 6.2$\pm$0.2 & 1.4$\pm$0.3 & 10.78$\pm$0.14 & HETG \\
\ion{Ne}{ix} & 0.67$\pm$0.04 & 6.6$\pm$0.1 & 2.9$\pm$0.3 & 10.95$^{+0.2}_{-0.5}$ & HETG \\
\ion{Mg}{xi} & 0.75$\pm$0.08 & 6.7$\pm$0.1 & 2.2$\pm$0.3 & 12.4$^{+0.2}_{-0.3}$ & HETG \\
\ion{Si}{xiii}& 0.66$\pm$0.05 & 6.7$\pm$0.1 & 3.9$\pm$0.5 & $\la$11.5 & HETG \\
\hline
\end{tabular}
\end{footnotesize}
\end{center}
\end{table}

\begin{table}
\caption{Electron densities calculated from \ion{Fe}{xxi} and
  \ion{Fe}{xxii} line ratios.}\label{tab:densFe}
\tabcolsep 3.pt
\begin{footnotesize}
\begin{center}
\begin{tabular}{llccl}
\hline \hline
No.$^a$ & Ion & Lines ratio & {$T_{\rm max}$} & {log $n_{\rm e}$(cm$^{-3}$)} \\
\hline
1 & \ion{Fe}{xxi} & $\lambda$12.284/$\lambda$12.499 & 7.0 & 12.80$\pm 0.13$  \\
5 & \ion{Fe}{xxi} & $\lambda$12.393/$\lambda$12.499 & 7.0 & 12.29$\pm 0.14$  \\
2 & \ion{Fe}{xxii} & $\lambda$11.770/$\lambda$11.932 & 7.1 & 13.22$\pm 0.12$  \\
3 & \ion{Fe}{xxii} & $\lambda$11.932/$\lambda$8.9748 & 7.1 & 12.0$^{+0.2}_{-0.5}$  \\
4 & \ion{Fe}{xxii} & $\lambda$11.932/$\lambda$11.802 & 7.1 & 12.4$^{+0.2}_{-0.3}$  \\
6 & \ion{Fe}{xxii} & $\lambda$11.490/$\lambda$11.932 & 7.1 & 12.74$\pm 0.14$  \\
7 & \ion{Fe}{xxii} & $\lambda$12.210/$\lambda$11.932 & 7.1 & 12.79$\pm 0.12$  \\
\hline
\end{tabular}
\end{center}
($a$): Number corresponding to labels in Fig.~\ref{denselecFe}
\end{footnotesize}
\end{table}

The steep increase of the plasma pressure with temperature is possibly
one of the most intriguing results provided by the currently available
Chandra high-resolution spectra of active stars \citep{fav03}.
A similar trend was recently found in Capella by \citet{argi03}, and
tentatively interpreted as due to the presence of different classes of coronal
loop structures in isobaric conditions, having increasing pressures but
decreasing volume filling factors for increasing maximum temperature of
the trapped plasma.

The effective scale sizes and volumes of the structures responsible for 
the X-ray emission from stellar coronae essentially depend on two
parameters: the plasma pressure scale height and the strength of the
magnetic field required for plasma confinement. In the case of AB~Dor,
having a stellar radius $R_* \sim 1$\,R$_{\sun}$ and a surface gravity
$g \sim 0.8$\,g$_{\sun}$ \citep{mag00}, we get pressure scale heights
$H_{\rm p} \sim 1.5 \times 10^{10}$\,cm ($0.2 R_*$) for coronal
loops with maximum temperature $T_{\rm max} = 2 \times 10^6$\,K,
and $H_{\rm p} > 7.5 \times 10^{10}$\,cm ($> 1.1 R_*$) for the structures
having $T_{\rm max} = 10^7$\,K or hotter, which represent the
dominant class in the corona of AB~Dor. If the ratio between the plasma
pressure and the magnetic field pressure, $\beta$, is larger than unity
at the maximum height, $L \le H_{\rm p}$, 
above surface dictated by the pressure scale
height, then magnetic confinement is effective at these heights and the
volume of the emitting plasma in the visible stellar hemisphere 
can be expressed as
\begin{equation}
V_{\rm max} \approx 2 \pi f R_*^2  L
\end{equation}
where $f$ is a surface filling factor. The minimum field
required to confine the plasma at a height $L$ is given by 
\begin{equation}
B_{\rm min}(L)= (8 \pi p_{\rm e})^{1/2}
\end{equation}
From the available data we derive  
$B_{\rm min} \sim 30$\,G at $T \sim 2 \times 10^6$\,K 
($p_{\rm e} = 35$\,dyn cm$^{-2}$) and
$B_{\rm min} \sim 460$\,G at $T \sim 10^7$\,K ($p_{\rm e} = 8.5 \times
10^3$\,dyn cm$^{-2}$). 
Donati \& Collier Cameron (1997) found that up to 20\% of the surface of 
AB~Dor is covered with magnetic regions having typical values of 500\,G 
and peaks of 1.5\,kG.
Assuming that magnetic field at the stellar surface is 1.5\,kG and that
the field strength
decreases with height as in a magnetic dipole ($B\sim  r^{-3}$) 
rooted at the base of the
convection zone (so from $r \sim 0.5 R_*$ to $r \sim 0.5 R_* + L$), 
we estimate that the plasma at $2 \times 10^6$\,K can be confined up
to heights $L = 9.4 \times 10^{10}$\,cm ($1.3 R_*$), while in the
worst case of the high temperature ($10^7$\,K) coronal structures, 
the maximum height is $L = 1.7 \times 10^{10}$\,cm ($0.24 R_*$).
Since the characteristic height of the high temperature structures is 
smaller than $H_{\rm p}$, the isobaricity of the plasma is ensured. 
We can estimate the filling factor $f$ by equating $n_{\rm e}^2
V_{\rm max} = EM$, 
where $n_{\rm e}$ and $EM$ are the plasma density and volume
emission measure derived from our analysis. In this way we derive
$f \sim 2 \times 10^{-6}$ at $T \sim 10^7$\,K. 
In the case of the low-temperature plasma, the observed emission
originates mostly from the plasma below the pressure scale height
$H_{\rm p}$, and the effective filling factor can be derived by
replacing $L$ with $H_{\rm p}$, resulting in $f \sim 2 \times 10^{-4}$
at $T \sim 2 \times 10^6$\,K.
These filling factors can be
usefully compared with the $\sim 7$\% fraction of surface in the polar cap of
AB~Dor above $60^{\degr}$ latitude: this is a region always in view as the
star rotates, where high field strengths are suggested by
Zeeman-Doppler images \citep{jar02} and where a non-potential
field component (i.e.\ capable of providing free energy to power flares) 
is possibly present \citep{hus02}.
Larger filling factors are possible if the loops are significantly
shorter than the stellar radius. 

A more refined model of the size, strength and orientation of the
magnetic regions is however required to constrain better the possible
sizes and locations of the X-ray emitting regions in the corona of
AB~Dor, as suggested by the simulations performed by \citet{jar02}. 
These authors have modeled the AB~Dor coronal X-ray emission
by extrapolating the magnetic field from the stellar surface to the
corona using as a basis Zeeman-Doppler maps and assuming the field to be
potential and the trapped plasma in hydrostatic equilibrium. However,
their results rely on the further assumption of an isothermal plasma,
which is clearly recognized as a too simplistic approximation.
Based on the above model, Jardine et al.\ conclude that in most of the
cases they have explored, the coronal emission of AB~Dor should exhibit
little rotational modulation: in fact, assuming low plasma densities, the
corona turns out to be very extended (in order to account for the observed
total volume emission measure) and hence the X-ray emission is little 
affected by the stellar rotation, while in the high-density case the
emitting corona is more compact, but the X-ray brightest regions are at
high latitudes and always visible as the star rotates. Exception to this
behavior is predicted by models having high-temperature ($T = 10^7$\,K)
plasma with densities in excess of $10^{12}$\,cm$^{-3}$, i.e.\ with exactly
the characteristics of the plasma near the peak of the emission measure
distribution, as derived in this work. The occurrence and amplitude of the 
rotational modulation of the X-ray emission from this hot plasma is an
issue that we intend to explore in the next step of our ongoing
investigation.

The corona of AB~Dor, even outside strong isolated flares, appears
quite variable on time scales shorter than the rotation period.
The only possible analogy with the case of the solar corona
if a behavior where small-scale flares are continuously occurring
in what can be defined a non-steady quiescent X-ray emission state.
However, the stability found in the
peak of the EMD at log~$T$(K)$\sim$6.9 suggests coronal structures in
stationary condition, and a simple increase in the number of loops having
maximum temperatures in the range spanned by the EMD plateau
may explain the variations of the EMD observed at the times of the
lowest and highest emission level observed up to now by XMM.

In conclusion, we summarize the main results of the present work as follows:

--  The Emission Measure Distribution (EMD) has been calculated for the
  plasma of AB~Dor by measuring the line fluxes in XMM/RGS and
  Chandra/HETG spectra, showing consistent results. The EMD is described
  by a quite stable structure, 
  with a steep increase ($EM[T] \propto T^\alpha$,
  with $\alpha =$ 4--5) up to the peak at log(T)=6.9, followed by a
  plateau in the range log(T)=6.9--7.3. The EMD during the highest
  and lowest X-ray emission levels shows an increment in the
  amount of material that is rather uniform at all temperatures.

-- Element abundances in the corona of AB~Dor follow an
  intermediate behavior  
  between the solar-like FIP (First Ionization Potential)
  effect, and the so called ``inverse FIP effect'' observed in other
  active stars. 

-- High electron densities were measured using He-like triplets
  and \ion{Fe}{xxi} and \ion{Fe}{xxii} line ratios. Together with the
  volume emission measure they allow to put constrains 
  on the surface filling factor of the emitting regions and on 
  the strength of magnetic fields required for plasma confinement.

-- The available data are consistent with a scenario of a corona composed 
  by several families of loops, shorter than but comparable to the stellar 
  radius and in isobaric conditions, having plasma pressures increasing 
  with the maximum plasma temperature; the surface filling factors of
  these structures is small.
  These structures can be easily accommodated in the stellar polar cap,
  where strong magnetic
  fields possibly in a non-potential state have been proposed.
  Larger filling factors are possible if the loops are significantly
  shorter than the stellar radius.


\begin{acknowledgements}
We would like to thank D. P. Huenemoerder and J. Houck (MIT/CXC) 
for their help in the use of ISIS, and its application to the analysis
of XMM data, and to N. Brickhouse for her help in the interpretation
of atomic data and line identifications.
We acknowledge support by the Marie Curie Fellowships Contract No.
HPMD-CT-2000-00013. 
We have made use of data obtained through the XMM-Newton Science Data
Archive, operated by ESA at VILSPA, and the Chandra Data Archive,
operated by the Smithsonian Astrophysical Observatory for NASA. This
research has also made use of NASA's Astrophysics Data System Abstract
Service.
\end{acknowledgements}

\end{document}